# Orbital Design of Flat Bands in Non-Line-Graph Lattices via Line-Graph Wavefunctions

Hang Liu,[1,2,3] Gurjyot Sethi,[1] Sheng Meng,[2,3,*] and Feng Liu[1,†]

[1] *Department of Materials Science and Engineering, University of Utah, Salt Lake City, Utah 84112, USA*

[2] *Songshan Lake Materials Laboratory, Dongguan, Guangdong 523808, People's Republic of China*

[3] *Beijing National Laboratory for Condensed Matter Physics and Institute of Physics, Chinese Academy of Sciences, Beijing 100190, People's Republic of China*

**Abstract:** Line-graph (LG) lattices are known for having flat bands (FBs) from the destructive interference of Bloch wavefunctions encoded in pure *lattice* symmetry. Here, we develop a generic atomic/molecular *orbital* design principle for FBs in non-LG lattices. Based on linear-combination-of-atomic-orbital (LCAO) theory, we demonstrate that the underlying wavefunction symmetry of FBs in a LG lattice can be transformed into the atomic/molecular orbital symmetry in a non-LG lattice. We illustrate such orbital-designed topological FBs in three 2D non-LG, square, trigonal, and hexagonal lattices, where the designed orbitals faithfully reproduce the corresponding lattice symmetries of checkerboard, Kagome, and diatomic-Kagome lattices, respectively. Interestingly, systematic design of FBs with a high Chern number is also achieved based on the same principle. Fundamentally our theory enriches the FB physics; practically it significantly expands the scope of FB materials, since most materials have multiple atomic/molecular orbitals at each lattice site, rather than a single *s* orbital mandated in graph theory and generic lattice models.

# I. INTRODUCTION

Electronic properties of crystals are generally determined by four fundamental degrees of order: lattice, orbital, charge, and spin [1]. One distinguished manifestation of lattice symmetry in electron band structure is topological flat band (TFB) in line-graph (LG) lattices [2-5]. In graph theory, a LG is made by connecting the centers of edges sharing a common vertex of a graph. It is proved that [6-8] the Laplacian operator of a LG is equivalent to the electronic Hamiltonian of the corresponding LG lattice, which has ubiquitously a constant eigenvalue, i.e., a FB. The topology of a FB hosted in a LG lattice can be assessed by the existence of singular band touching point with a dispersive band at a high-symmetry $k$ point [9-12], differing from an isolated trivial FB, such as the one in Tasaki lattices [13-15]. When the degeneracy of the touching point is lifted, the gapped TFB has a nonzero Chern number [9]. Due to its quenched kinetic energy and nontrivial topology, there exists a rich spectrum of physics associated with TFB, such as ferromagnetism [13,16,17], superconductivity [18-20], Wigner crystallization [21-23], fractional quantum Hall effect [24-27], Weyl fermion [28], and excitonic insulator [29]. Recent discovery of superconductivity in twisted bilayer graphene has further boosted the interest in FBs [30-33].

The lattice, orbital, charge, and spin degrees of order are interdependent with each other. Of particular interest here is the transformation between lattice and orbital symmetry. Some generic *lattice* models (namely one *s*-orbital per site), based on LG [2-8,34-46], cell [13-15,47-49], and compact localized state (CLS) construction [10,47,50], have been developed for FBs [see Section I of Supplemental Material (SM) [51]]. Also, a couple of specific models have been shown for FBs in non-LG lattices [22,35,52], such as the hexagonal lattice with ($p_x$, $p_y$) orbitals [22,53-55]. However, a generic *orbital* model for TFB construction, including high-Chern-number FB, is still lacking, which is important and useful since real materials usually consist of multiple atomic/molecular orbitals on each lattice site. In general, our understanding of fundamental relationship between lattice and orbital symmetry regarding FBs is far from complete.

In this work, we introduce a generic orbital design principle for TFBs, based on linear-combination-of-atomic-orbitals (LCAO) theory, which transforms the symmetry of lattice wavefunctions in LG lattice into "molecular" orbital (MO) symmetry in non-LG lattice. Applying this principle, we predict new FB lattice/orbital systems and explain the few existing ones. It also enables a systematic orbital design of FBs with a high Chern number in various lattices. Using the tight-binding method, we calculate the band structures of three most common non-LG, square, trigonal, and hexagonal lattices, by employing the combinations of orbitals that are symmetry-transformed from a subset of lattice wavefunctions of the LG, checkerboard, Kagome, and diatomic-Kagome lattices, respectively. We are able to produce TFBs in these lattices with *all* the possible orbital combinations, as summarized in Table I at the end, much beyond a few specific cases found previously by physical intuition.

## II. COMPUTATIONAL METHODS

For tight-binding model calculations, we employed the well-known two-center bond integrals initially derived by J. C. Slater and G. F. Koster [56]. All moment-space Hamiltonians without spin-orbit coupling (SOC) can be found in Section IV of SM [51]. Furthermore, onsite SOC (Table S1 [51,57]) is considered to break the degeneracy of singular touching points between the flat and dispersive band. For the two-dimensional systems in this work, the orbital bases with separate spin-up and -down channels $\{|o_1,\uparrow\rangle, \ldots, |o_n,\uparrow\rangle; |o_1,\downarrow\rangle, \ldots, |o_n,\downarrow\rangle\}$ are used, so the onsite-SOC contribution to Hamiltonian is written as $\lambda \mathbf{L} \cdot \mathbf{S} = \frac{\lambda}{2}\begin{bmatrix} L_z & 0 \\ 0 & -L_z \end{bmatrix}$ [58] with SOC strength $\lambda$, orbital angular momentum $\mathbf{L}$, and spin angular momentum $\mathbf{S}$, in which the matrix $L_z$ is derived based on Supplementary Table 1. The spin $z$ component is not mixed by the onsite SOC, manifesting $z$ component is still a good quantum number. For the spin-polarized band indexed as $n$, Chern invariant $C_n = \frac{1}{2\pi}\int_{BZ} d^2k\, \Omega_n$ is calculated by integrating Berry curvature in the FBZ [59]. The momentum-space Berry curvature is $\Omega_n(\mathbf{k}) = -\sum_{n'\neq n} \frac{2Im\langle\psi_{n\mathbf{k}}|\hat{v}_x|\psi_{n'\mathbf{k}}\rangle\langle\psi_{n'\mathbf{k}}|\hat{v}_y|\psi_{n\mathbf{k}}\rangle}{(E_{n'\mathbf{k}}-E_{n\mathbf{k}})^2}$, where $\hat{v}_x$ and $\hat{v}_y$ are velocity operators along the $x$ and $y$ directions. The Chern invariant is calculated for spin-up channel in this work.

## III. RESULTS AND DISCUSSION

### A. General formulation of orbital-design principle.

We first discuss a general formulation of the orbital designed TFBs in non-LG lattices. Consider a LG lattice consisting of $n$ sites per unit cell, such as $n = 2$ in a checkerboard lattice (LG of square lattice) in Fig. 1 (grey dots), with one $s$-orbital per site ($\varphi_s$). Let us partition the LG by grouping $m$ sites with labels ($A$, $B$, $C$, $\cdots$) together as a "molecule" (periodically repeated), the resulting MOs are constructed from LCAO theory as

$$\emptyset_{MO} = \sum_{i=A,B,C,\cdots}^{m} c_i \varphi_{is}, \tag{1}$$

and treat the center of this molecule as one site with $n$ MOs in a new lattice, which will generally be a non-LG lattice. For example, in Fig. 1(a), we chose $m = n = 2$, then the new lattice is a square lattice with two MOs on each site (blue dots). This operation transforms the lattice symmetry of a LG checkerboard lattice into the orbital symmetry of a non-LG square lattice. Importantly both lattices must have the same band structure, including the FB, because they have the equivalent Hamiltonian with only different basis expansions for the Bloch wavefunctions in the same lattice partition, the former expanded in single $s$-orbitals and the latter in multi-MOs. The *symmetry* (or type) of the MOs is determined by coefficients $c_i$ in Eq. (1), in particular the sign of $c_i$ on the $m$ LG lattice sites, which can be obtained from the nodes of Bloch wavefunctions at the $\Gamma$ point. Note that the basis transformation is independent of $\boldsymbol{k}$ points; in other words, the Bloch states at every $\boldsymbol{k}$ point are solved with the same $s$-orbital (MO) basis in the LG (non-LG) lattice. For example, the two MOs in Fig. 1(a) have the general form of $\emptyset_{MO}^{1,2} = (\varphi_{As} \pm \varphi_{Bs})/\sqrt{2}$ [see calculation results in Fig. 2(a)], indicating one $s$- and one $p$-orbital. This is different from the $k$-resolved symmetry analysis of the whole-lattice Bloch states, used to assess band topology [60,61].

Also, different MOs may result from partitioning different number of $m$ sites. For example, in Fig. 1(b), we chose $m = 2n = 4$, the new lattice is again a square lattice, but the two MOs on each site (blue dots) have $s$- and $d$-symmetry, respectively [see calculation results in Fig. 2(e)]. In doing so, the new square lattice has a supercell size twice as large as the original checkerboard lattice, as the two MOs are transformed from four *s*-orbitals. This indicates band folding must accompany with this basis transformation, since the band structure is independent of basis representations. Interestingly, this would lead to multiple "folded" singular band touching points between the flat and dispersive band, indicating the FB with a high Chern number. Here we consider the Chern FB defined for one spin component (see Section II).

**B. Square lattice.**

Next, we illustrate the above design principle by tight-binding band calculations of specific examples. The checkerboard lattice, having site A and B in a unit cell, hosts a FB touched with a dispersive band (Fig. S1 in SM [51]). In Fig. 2(a), we plot the two eigenstates at $\Gamma$, $\psi_{E=-3t} = \frac{1}{\sqrt{2}}(|A\rangle + |B\rangle)$ and $\psi_{E=t} = \frac{1}{\sqrt{2}}(|A\rangle - |B\rangle)$, centered at the middle of A and B. They may be viewed as two MOs sitting on the same site in a square lattice, one with $s \sim |A\rangle + |B\rangle$ and the other with $p \sim |A\rangle - |B\rangle$ symmetry polarized along the diagonal direction, as shown in Fig. 2(a,b). The band structure of this square lattice is calculated, as shown in Fig. 2(c), using the following nearest-neighbor (NN) and next-NN (NNN) hopping integrals (note: the two-center Slater-Koster integrals [56,62] is scaled by a common factor *t*.)

$$t_{ss\sigma}^{NN} = -t_{sp\sigma}^{NN} = -\frac{1}{2}, t_{pp\sigma}^{NN} = -t_{pp\pi}^{NN}, t_{ss\sigma}^{NNN} = -\frac{1}{4}, t_{sp\sigma}^{NNN} = \frac{1}{2\sqrt{2}}, t_{pp\sigma}^{NNN} = \frac{1}{2}. \tag{2}$$

One may also partition the checkerboard lattice wavefunctions in Fig. 2(a) differently, by grouping four instead of two lattice sites, as illustrated in Fig. 2(e). Expanding the $\Gamma$-point lattice wavefunctions into a four-site basis in a $\sqrt{2} \times \sqrt{2}$ checkerboard superlattice gives rise to two MOs corresponding to $\psi_{E=-3t} = \frac{1}{2}(|A_1\rangle + |A_2\rangle + |B_1\rangle + |B_2\rangle)$, and $\psi_{E=t} = \frac{1}{2}(|A_1\rangle + |A_2\rangle - |B_1\rangle - |B_2\rangle)$, with *s* and *d* symmetry in a square lattice [Fig. 2(e,f)]. This enables an alternative design of FB in a square lattice, and Figure 2(g) shows the resulting band structure calculated using (*s*, $d_{xy}$)-hopping integrals of $t_{ss\sigma}^{NN} = -t_{dd\pi}^{NN} = -\frac{1}{2}, t_{ss\sigma}^{NNN} = -\frac{1}{4}, t_{sd\sigma}^{NNN} = -\frac{1}{2\sqrt{3}}$, and $t_{dd\sigma}^{NNN} = -\frac{1}{3}$.

The band structure in Fig. 2(g) is different but related to that in Fig. 2(c) by band folding.

Specifically, the bands in Fig. 2(g), calculated from the 1 × 1 unit cell [solid square in Fig. 2(f)] with a "1 × 1" first Brillouin zone (FBZ) [solid square of inset in Fig. 2(g)], can be folded into the bands of a $\sqrt{2} \times \sqrt{2}$ cell [dashed rhombus in Fig. 2(f)] with a "$\frac{1}{\sqrt{2}} \times \frac{1}{\sqrt{2}}$" FBZ [dashed rhombus of inset in Fig. 2(g)], with the M, middle point of Γ-M, and X of the former folded into the Γ, X, and M of the latter, respectively. The folding produces two sets of degenerate checkerboard bands (Fig. S2) having the identical band dispersions as in Fig. 2(c). Interestingly, this renders the FB in Fig. 2(g) to have a Chern number of −2 for one spin channel, as manifested by the observation of FB touched with the dispersive band at two X points within the FBZ, as explained below.

At each band touching point, the Berry curvature of singular Bloch wavefunctions of TFB diverges, in association with the $N-1$ CLSs for a finite lattice with $N$ sites, and 2 topological noncontractible loop states (NLSs), i.e., extended boundary states, in real space [9-12]. In LG lattices, the CLS is resulted from destructive interference (phase cancellation) of lattice hopping induced solely by lattice symmetry, as reflected by the alternating nodal signs of wavefunction on an even-edged plaquette, e.g., a rhombus in a checkerboard lattice (Fig. S1). The CLS of orbital-designed FBs is more complex, as illustrated in Fig. 2(d). The Bloch state of FB in Fig. 2(c) is calculated as $\psi_{\mathbf{k}}^{FB} = i\frac{1}{\sqrt{2}}\sin\frac{k_1+k_2}{2}|s\rangle + \cos\frac{k_1}{2}\cos\frac{k_2}{2}|p\rangle$ with $k_j = \mathbf{k} \cdot \mathbf{a}_j$ ($\mathbf{a}_j$, lattice vector; $j$ = 1, 2), whose Fourier transformation $\psi_{\mathbf{R}}^{FB} = \int_{BZ} d\mathbf{k} e^{-i\mathbf{k}\cdot\mathbf{R}} \psi_{\mathbf{k}}^{FB}$ produces a real-space CLS on a square plaquette centered at $\mathbf{R}$ [Fig. 2(d)]. It consists of nodal wavefunctions of $\frac{|p\rangle}{4}, -\frac{|s\rangle}{2\sqrt{2}} + \frac{|p\rangle}{4}, \frac{|p\rangle}{4}$, and $\frac{|s\rangle}{2\sqrt{2}} + \frac{|p\rangle}{4}$ at four vertices of the plaquette, respectively. Electron hoppings outward from the CLS to its surrounding lattice sites are completely forbidden, which can be shown by analyzing hoppings based on Eq. (2). For example, the hoppings to the site above site 1 come from site 1 and 2 in Fig. 2(d), which are respectively $\frac{1}{4}\left[\frac{1}{2}(t_{pp\sigma}^{NN} + t_{pp\pi}^{NN}) - \frac{1}{\sqrt{2}}t_{sp\sigma}^{NN}\right] = -\frac{t}{8\sqrt{2}}$ and $-\frac{1}{2\sqrt{2}}t_{ss\sigma}^{NNN} + \frac{1}{4}t_{pp\pi}^{NNN} = \frac{t}{8\sqrt{2}}$, and cancels out with each other. The same is true for all other hoppings. Likewise, the FB in sd-orbital model supports a CLS on a square plaquette with linearly combined s and d orbitals on its vertices [Fig. 2(h)], whose outward hoppings also vanish (see details in Section II of SM [51]).

Thus, the orbital symmetry in a non-LG lattice plays the role of lattice symmetry in a LG lattice in conditioning the destructive interference of Bloch states to form a CLS. Besides the localized CLSs, the extended FB NLSs exist (see details in Section III of SM [51]), indicating the nontrivial topology of the orbital-designed FBs [9-12].

In real materials, there are usually multiple atomic/molecular orbitals on each lattice site contributing to the bands near Fermi level. In the proposed lattice-orbital transformation, the number of sites in the LG lattice equals the number of orbitals in the non-LG lattice per unit- or super-cell. So, to design FB with more than two orbitals in a square lattice, one may find another LG lattice, instead of checkerboard lattice, with more sites per cell. One such choice is the diamond-octagon (diatomic-checkerboard) lattice, the LG of Lieb lattice [43]. This leads to dual TFBs in a square lattice. Without losing generality, let us first choose three orbitals ($s$, $p_x$, $p_y$) [Fig. 3(a,b)], and the following NN hopping integrals

$$t_{ss\sigma} = -\frac{1}{8}, t_{pp\sigma} = \frac{1}{4}, t_{sp\sigma} = \frac{1}{4\sqrt{2}}. \tag{3}$$

The resulting band structure consists of two FBs touched with one dispersive band in between [Fig. 3(c)]. They correspond exactly to the top three bands of diamond-octagon lattice (Fig. S3), whose lattice wavefunctions at $\Gamma$ indeed display the $s$, $p_x$, and $p_y$ orbital symmetry, respectively, as shown in Fig. 3(a), following our design principle.

The CLS is analyzed to understand how the kinetic energy of the dual TFBs is quenched. As illustrated in Fig. 3(d), the lower FB supports a CLS on a square plaquette with bonding nodal wavefunctions $|s\rangle + \frac{1}{\sqrt{2}}(|p_x\rangle - |p_y\rangle)$, $|s\rangle - \frac{1}{\sqrt{2}}(|p_x\rangle + |p_y\rangle)$, $|s\rangle + \frac{1}{\sqrt{2}}(-|p_x\rangle + |p_y\rangle)$, and $|s\rangle + \frac{1}{\sqrt{2}}(|p_x\rangle + |p_y\rangle)$ at four vertices, respectively; while the upper-FB CLS consists of four anti-bonding vertex states. Based on Eq. (3), both CLSs have vanishing outward hoppings. For example, the one from site 1 to the site above is $(t_{ss\sigma} + t_{sp\sigma}) - \frac{1}{\sqrt{2}}(-t_{sp\sigma} + t_{pp\sigma}) = 0$ (see others and topological NLSs in Section II and III of SM [51], respectively). It once again confirms that the orbital symmetry underlines the destructive interference of Bloch wavefunctions for our theoretically designed TFBs.

The diamond-octagon lattice has total four sites per unit cell and hence four bands; the wavefunction of the fourth bottom isolated band has the $d_{x^2-y^2}$ symmetry (Fig. S3). It can be shown that by changing the sign of lattice hopping integrals, the position of $s$- and $d_{x^2-y^2}$-band is interchangeable. Consequently, an alternative design of the top three bands with dual FBs is to use ($d_{x^2-y^2}$, $p_x$, $p_y$) in place of ($s$, $p_x$, $p_y$) orbitals in a square lattice (Fig. S4), as found previously [35]. One may also include four orbitals ($s$, $p_x$, $p_y$, $d_{x^2-y^2}$) in a square lattice, which produces two sets of checkerboard bands of opposite chirality (Fig. S5). They are dubbed as Yin-Yang checkerboard bands, in analogy to the Yin-Yang Kagome bands [41]. The above hopping integrals in square lattice produce a perfect FB, same as in the corresponding LG lattice with ideal hopping integrals. Ideally, specific inter-orbital hopping integrals are assumed in the non-LG lattices to produce FBs with perfect flatness, same as in the corresponding LG lattice with ideal hopping integrals. Usually, a small deviation from ideal hopping conditions in either LG or non-LG lattices leads to finite dispersion of FBs, but without changing the physics qualitatively ([64], Fig. S6).

In the present study, we focus on the LG of bipartite lattices whose FB is singularly touched with a dispersive band, and hence topologically nontrivial [9-12], while the LG of non-bipartite lattices have an isolated FB that is topologically fragile [4,8,36]. Each singular band touching point can be viewed as a Berry flux center, in analogy to Dirac/Weyl point [9,63], contributing to one integer Chern number of ±1. This can be clearly illustrated by evaluating evolution of band structure and Berry curvature $\Omega$ of FB as a function of increasing SOC strength, as shown in Fig. 4 for the case of $sp^2$ square lattice as an example. One sees that with a diminishing SOC [Fig. 4(a)] towards zero, $\Omega$ vanishes everywhere except for around the Γ point where it diverges going to infinity on a tiny small circle. With the increasing SOC, the distribution of $\Omega$ gradually broadens around Γ on a band of ring. In all cases, integration of gives a Chern number of 1. Indeed, this is confirmed by adding SOC to open a gap, and calculating the FB Chern number in all the lattices considered (Fig. S7). Each singular band touching point contributes a Chern number of +1 or -1; therefore, the proposed orbital design provides also an effective way to realize high-Chern-number

FBs by introducing multiple band touching points as shown in Fig. 2(g) [see also Fig. 5(f) below].

**C. Trigonal lattice.**

Next, we demonstrate existence of TFB in a trigonal lattice, by orbital design from a Kagome lattice (LG of hexagonal lattice). Kagome lattices has three Γ-point eigenstates of $\frac{1}{\sqrt{3}}(|A\rangle + |B\rangle + |C\rangle)$, $\frac{1}{\sqrt{2}}(-|A\rangle + |B\rangle)$, and $\frac{1}{\sqrt{6}}(-|A\rangle - |B\rangle + 2|C\rangle)$ at $E = -4t$, $2t$, and $2t$ [Fig. 5(a)], which have the $s$, $p_x$, and $p_y$ orbital symmetry, respectively. So, using three orbitals ($s$, $p_x$, $p_y$) on each site [Fig. 5(b)] and the following NN hopping integrals

$$t_{ss\sigma} = -\frac{2}{3},\ t_{pp\sigma} = 1,\ t_{pp\pi} = -\frac{1}{3},\ t_{sp\sigma} = \sqrt{\frac{2}{3}}, \tag{4}$$

a FB appears to touch with two Dirac bands [Fig. 5(c)], similar to Kagome bands (Fig. S8). Since the lattice-orbital transformation has three lattice sites transformed into three orbitals in the same unit cell, there is no band folding.

Similar to the design of high-Chern-number FB in square lattice [Fig. 2(g) vs. 2(c)], instead of three sites (nodal points), one may consider the symmetry of Kagome lattice wavefunctions partitioned on twelve nodal points in a 2 × 2 supercell [Fig. 5(d)], leading to three ($s$, $d_{xy}$, $d_{x^2-y^2}$) orbitals in a trigonal lattice. Correspondingly, using ($s$, $d_{xy}$, $d_{x^2-y^2}$) basis in a 1 × 1 trigonal unit cell [Fig. 5(e)] and NN hopping integrals $t_{ss\sigma} = -\frac{2}{3}$, $t_{dd\sigma} = -\frac{4}{9}$, $t_{dd\pi} = 1$, $t_{sd\sigma} = -\sqrt{\frac{8}{27}}$, we obtain a band structure exhibiting a FB touched with Dirac bands [Fig. 5(f)], but different from the apparent looking of Kagome bands (Fig. S8). This is because they are unfolded from four sets of degenerate Kagome bands appearing as if in a "2 × 2" supercell with a "$\frac{1}{2} \times \frac{1}{2}$" FBZ [four dashed hexagons in the inset of Fig. 5(f)], with the Γ, K and M of the former unfolded from the Γ, K and Γ of the latter, respectively. Accordingly, there are four singular band touching points, one at Γ and three at M in the FBZ, with the former contributing a Chern number of +1 and the latter −3, adding to a net FB Chern number of −2 (Fig. S7). The FB CLS in a trigonal lattice is on a hexagon plaquette, with outward hopping all canceled out by orbital symmetry (Fig. S9).

**D. Hexagonal lattice.**

Lastly, we discuss the design of FBs in a hexagonal lattice with site A and B, where an already-known orbital basis is ($p_x$, $p_y$) with the hopping integral $t_{pp\sigma} = \frac{2}{3}$ [22]. The four Γ-point eigenstates are $-|A:p_x\rangle + |B:p_x\rangle$, $|A:p_y\rangle - |B:p_y\rangle$, $|A:p_x\rangle + |B:p_x\rangle$, and $|A:p_y\rangle + |B:p_y\rangle$, which have the same symmetry of diatomic-Kagome lattice wavefunctions (Fig. S10, S11), noticing that the diatomic-Kagome lattice is a generalized LG (i.e., two copies of LG) of hexagonal lattice. So, it again conforms to our generic orbital-design principle.

Furthermore, other orbital bases for TFBs in hexagonal lattice can be designed. As shown in Fig. 6(a,b), using ($d_{xy}$, $d_{x^2-y^2}$) orbitals on each site, and a NN hopping integral $t_{dd\sigma} = -\frac{8}{9}$, two FBs sandwiching two Dirac bands are obtained [Fig. 6(c)]. The doubly degenerate Γ-point eigenstates are $-|A:d_{xy}\rangle - |B:d_{xy}\rangle$ and $|A:d_{x^2-y^2}\rangle + |B:d_{x^2-y^2}\rangle$ ($|A:d_{xy}\rangle - |B:d_{xy}\rangle$ and $|A:d_{x^2-y^2}\rangle - |B:d_{x^2-y^2}\rangle$), respectively, having the same symmetries as those of diatomic-Kagome lattice Γ-point wavefunctions in Fig. 6(a). Adding another s orbital leads to Yin-Yang Kagome bands (Fig. S12), as found previously [41].

**IV. CONCLUSION**

We have developed a generic orbital design principle for FBs, including high-Chern-number FBs, in non-LG lattices via LG lattice wavefunctions, as summarized in Table I, for *all* possible orbital combinations. Importantly, the required hopping conditions are generally achievable in real materials, and some of the proposed designs have already been shown in real materials [41,52-55,64,65]. Also, more strict hopping conditions can be met by designing artificial lattice systems, where photonic and phonic FBs can be created in the non-LG lattices. Generally, the implementation of the design principle is more flexible with molecular than atomic orbitals. For example, the frontier MOs of a triangular graphene flake have the $p_z$, ($p_x$, $p_y$) and (s, $p_x$, $p_y$) symmetry, respectively, with a side length of 2, 3, and 4 benzene rings [41]. One may also generalize the design principle to three dimensional lattices.

## Acknowledgements

H.L. and S.M. thank financial support from the National Natural Science Foundation of China (Grants No. 12025407, 11934003, and 12147136), Chinese Academy of Sciences (Grant No. XDB330301), China Postdoctoral Science Foundation (Grant No. 2021M700163), and Guangdong Basic and Applied Basic Research Foundation (Grant No. 2021A1515110466). G.S. and F.L. are supported by U.S. DOE-BES (Grant No. DE-FG02-04ER46148).

---

[†] fliu@eng.utah.edu
[*] smeng@iphy.ac.cn

**Figures and Tables**

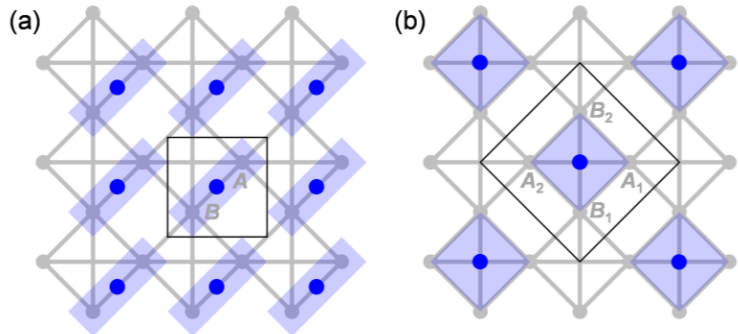

FIG. 1. (a) Illustration of lattice wavefunction symmetries, viewed on two sites (A and B, shaded) in a $1 \times 1$ checkerboard primitive cell (black thin lines), to be transformed into two orbitals on one site (blue dot) in a square lattice: $|s\rangle \sim |A\rangle + |B\rangle, |p\rangle \sim |A\rangle - |B\rangle$. (b) Same as (a) but viewed on four sites ($A_1$, $A_2$, $B_1$, and $B_2$, shaded) in a $\sqrt{2} \times \sqrt{2}$ checkerboard supercell (black thin lines). The two orbitals become $|s\rangle \sim |A_1\rangle + |A_2\rangle + |B_1\rangle + |B_2\rangle, |p\rangle \sim |A_1\rangle + |A_2\rangle - |B_1\rangle - |B_2\rangle$.

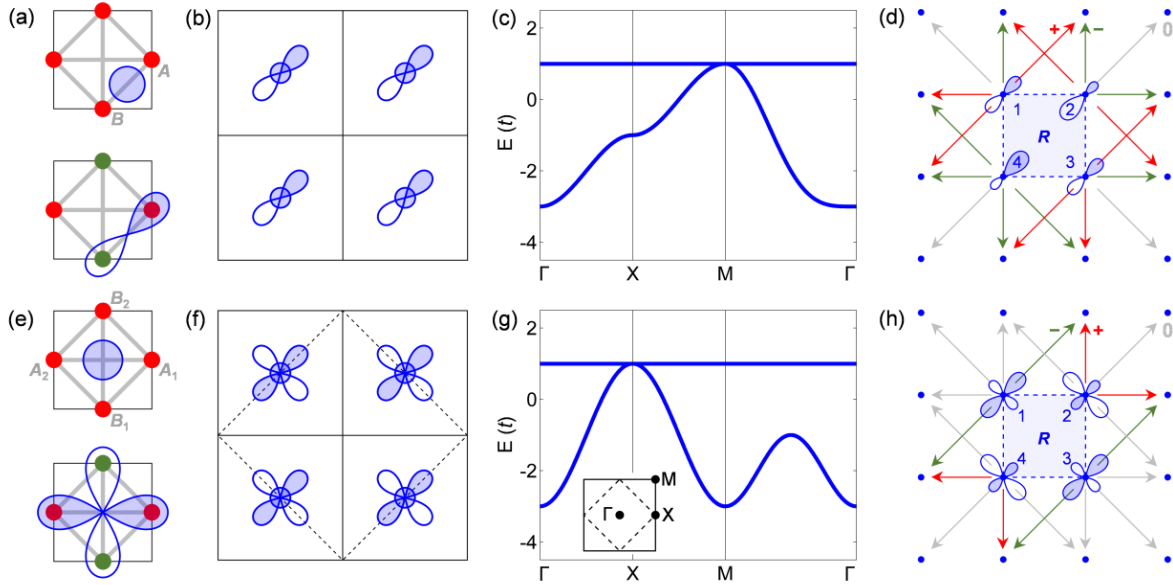

FIG. 2. TFB in a square lattice. (a) Γ-point wavefunctions of a checkerboard lattice (gray lines), exhibiting $s$ (upper) and $p$ orbital symmetry (lower). Red and green dots represent respectively positive and negative wavefunction nodes. (b) Square lattice with ($s$, $p$) orbitals. (c) Band structure of (b) with hopping integrals in Eq. (2). (d) The CLS of FB in (b) on a square plaquette, illustrating overall zero outward hoppings. Red, green, and gray arrows represent positive, negative, and zero hopping integrals, respectively. (e)-(h) Same as (a)-(d) with ($s$, $d_{xy}$) orbitals. The bands in (g) can be viewed as folded from (c), the inset of (g) shows the FBZs before (solid line) and after folding (dashed line).

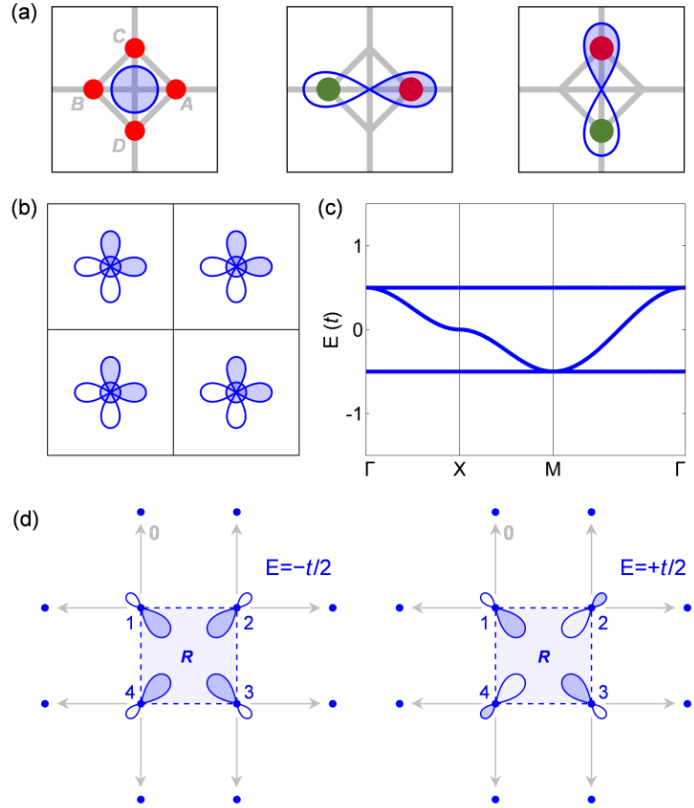

FIG. 3. Dual TFBs in a square lattice. (a) Three Γ-point wavefunctions of a diamond-octagon lattice, with $s$ (left), $p_x$ (middle), and $p_y$ orbital symmetry (right), respectively. (b) Square lattice with ($s$, $p_x$, $p_y$) orbitals. (c) Band structure of (b) with NN hopping integrals in Eq. (3). (d) The CLS of lower (left) and upper FB (right) in (c) on a square plaquette. Gray arrows indicate vanishing outward hoppings.

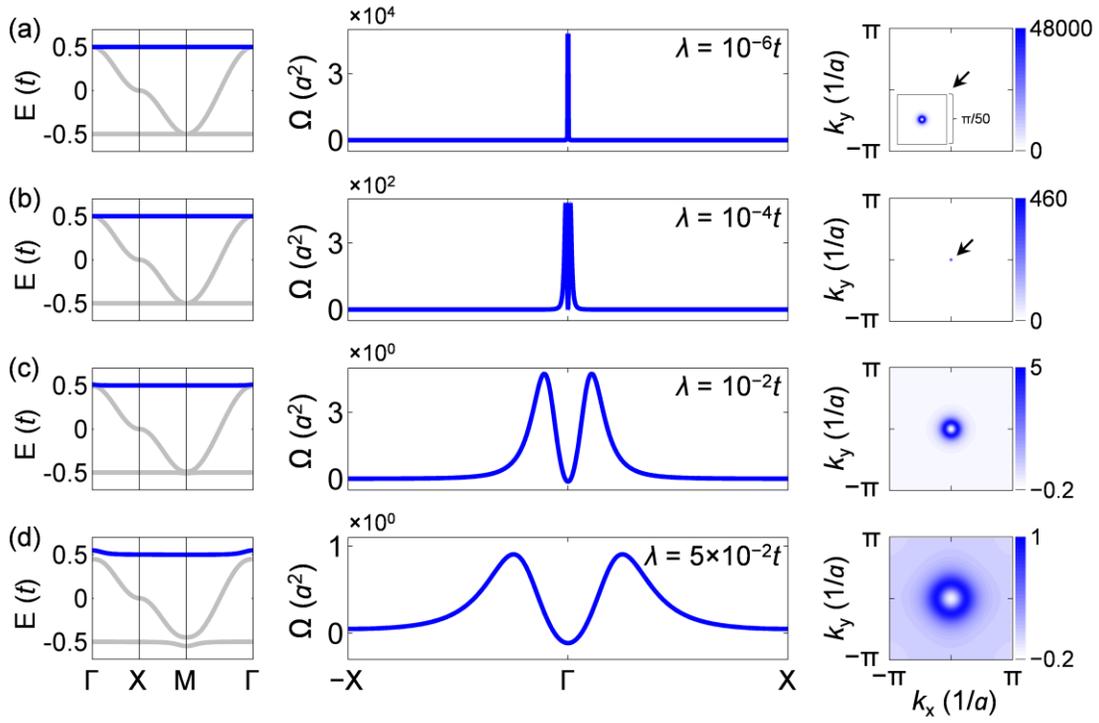

FIG. 4. SOC induced evolution of band structure and Berry curvature $\Omega$ of TFB in $sp^2$ square lattice with hopping parameters of Eq. (3). The strength of onsite SOC is (a) $\lambda = 10^{-6}t$, the inset shows a 100× magnification of dot at $\Gamma$, (b) $\lambda = 10^{-4}t$, (c) $\lambda = 10^{-2}t$, (d) $\lambda = 5 \times 10^{-2}t$. The calculated $\Omega$ is for upper TFB, in units of $a^2$ ($a$ is lattice constant).

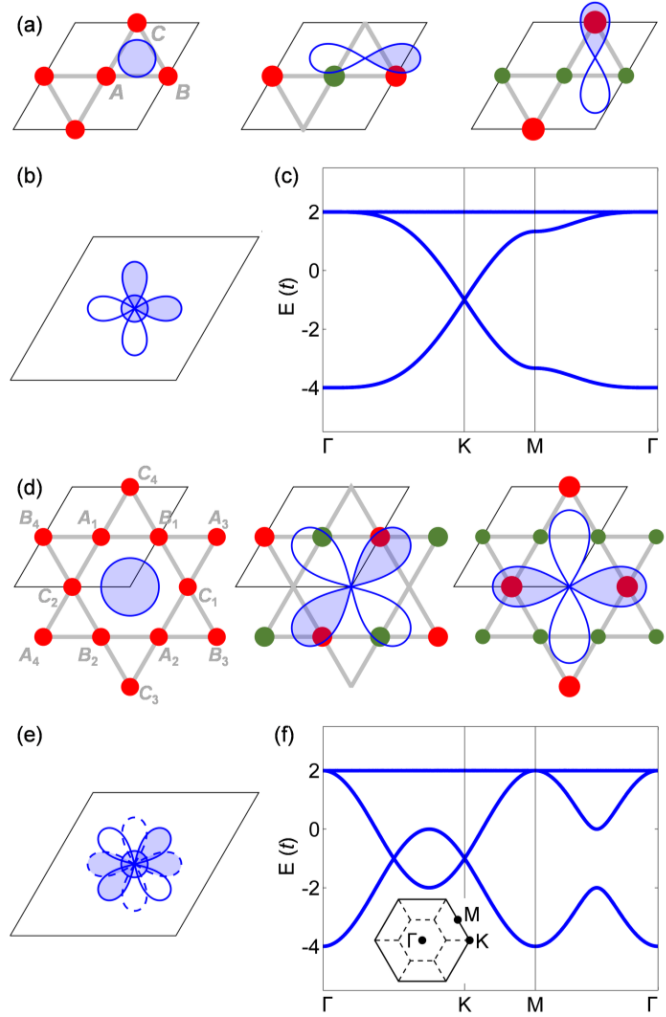

FIG. 5. TFB in a trigonal lattice. (a) Three Γ-point wavefunctions of a Kagome lattice, with $s$ (left), $p_x$ (middle), and $p_y$ orbital symmetry (right), respectively. (b) Trigonal lattice with ($s, p_x, p_y$) orbitals. (c) Band structure of (b) with NN hopping integrals in Eq. (4). (d-f) Same as (a-c) with ($s, d_{xy}, d_{x^2-y^2}$) orbitals. The bands in (f) can be viewed as folded from (c), the inset of (f) shows the FBZs before and after folding.

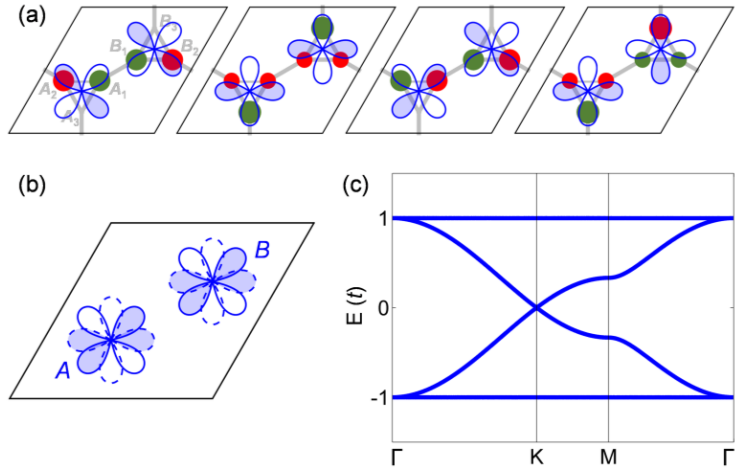

FIG. 6. Dual TFBs in a hexagonal lattice. (a) Four $\Gamma$-point wavefunctions of a diatomic-Kagome lattice, all with $d$ orbital symmetry. (b) Hexagonal lattice with ($d_{xy}$, $d_{x^2-y^2}$) orbitals on site A and B. (c) Band structure of (b) with a NN hopping integral $t_{dd\sigma} = -\frac{8}{9}$.

Table. I. Transformation from LG lattices to non-LG orbitals.

| LG lattice | Non-LG orbital |
|---|---|
| Checkerboard | Square ($s/d_{z^2}$, $p$), ($s/d_{z^2}$, $d$) |
| Diamond-octagon | Square ($s/d_{z^2}$, $p_x$, $p_y$), ($d_{x^2-y^2}$, $p_x$, $p_y$) |
| Kagome | Trigonal ($s/d_{z^2}$, $p_x$, $p_y$), ($s/d_{z^2}$, $d_{xy}$, $d_{x^2-y^2}$) |
| Diatomic-Kagome | Hexagonal ($p_x$, $p_y$), ($d_{xy}$, $d_{x^2-y^2}$) |

Supplemental Material for

# Orbital Design of Flat Bands in Non-Line-Graph Lattices via Line-Graph Wavefunctions

Hang Liu,[1,2,3] Gurjyot Sethi,[1] Sheng Meng,[2,3,*] and Feng Liu[1,†]

[1] *Department of Materials Science and Engineering, University of Utah, Salt Lake City, Utah 84112, USA*

[2] *Songshan Lake Materials Laboratory, Dongguan, Guangdong 523808, People's Republic of China*

[3] *Beijing National Laboratory for Condensed Matter Physics and Institute of Physics, Chinese Academy of Sciences, Beijing 100190, People's Republic of China*

The material contains following contents:

- **Supplemental Figures and Tables.**

  FIG. S1-S18.
  Table S1.

- **Supplemental Text.**

  I. Lattice construction of FBs.
  II. Compact localized state (CLS) of orbital-designed TFBs.
  III. Noncontractible loop state (NLS) of orbital-designed TFBs.
  IV. Momentum-space Hamiltonians without SOC.

___________________

[†]fliu@eng.utah.edu

[*]smeng@iphy.ac.cn

## Supplemental Figures

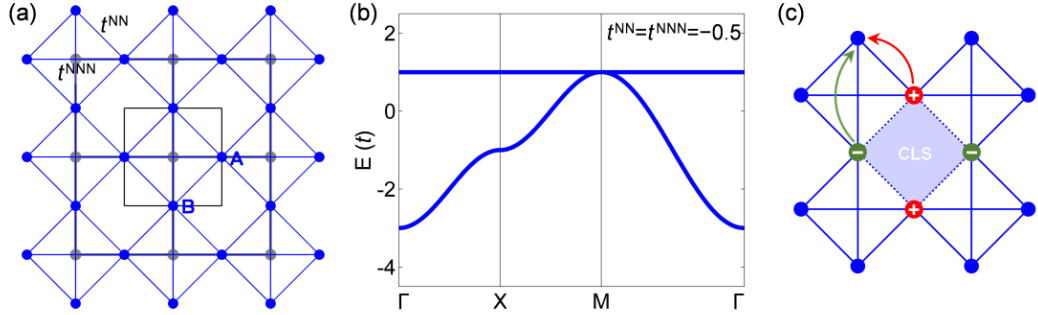

FIG. S1. TFB in a checkerboard lattice. (a) Illustration of checkerboard lattice (blue dots and lines) as LG of square lattice (gray dots and lines) with two sites (A, B) in the unit cell (black thin lines). (b) The band structure having one FB touched with one dispersive band at M point, calculated with $t^{NN} = t^{NNN}$. (c) Illustration of the compact localized state (CLS) of checkerboard FB on a rhombus plaquette (shaded), where red and green dots represent respectively positive and negative nodes of real-space wavefunction on the four vertices of plaquette, leading to canceling outwards hoppings (arrows), with the condition $t^{NN} = t^{NNN}$.

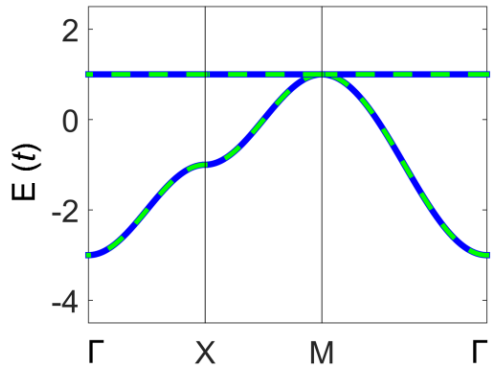

FIG. S2. Band structure of the $(s, d_{xy})$-orbital square lattice with a $\sqrt{2} \times \sqrt{2}$ supercell. The bands consist of two sets of degenerate checkerboard bands (blue solid and green dashed lines) as if folded from $1 \times 1$ primitive cell.

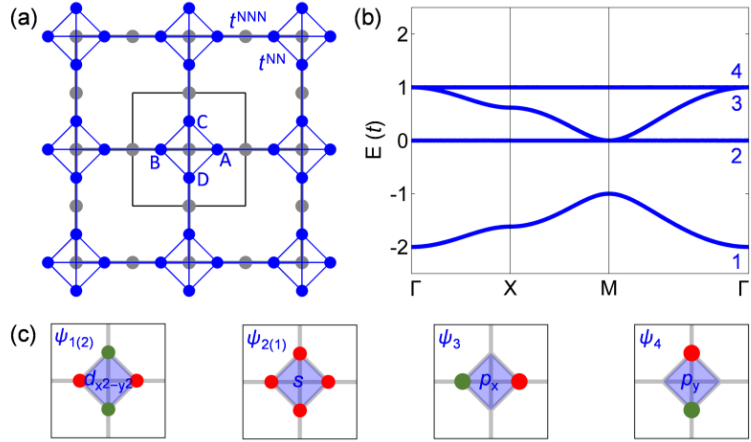

FIG. S3. TFBs in a diamond-octagon lattice. (a) Illustration of diamond-octagon lattice (blue dots and lines) as LG of Lieb lattice (gray dots and lines) with four sites (A, B, C, and D) in the unit cell (black thin lines). $t^{NN}$ and $t^{NNN}$ represent NN and NNN hopping integrals, respectively. (b) Band structure of (a) with $t^{NN} = -t^{NNN} = 0.5$ (or $t^{NN} = t^{NNN} = -0.5$). (c) Illustration of Γ-point lattice wavefunctions of the four bands in (b). Red and green dots represent positive and negative wavefunction nodes on the lattice sites, respectively, to indicate the corresponding orbital symmetry.

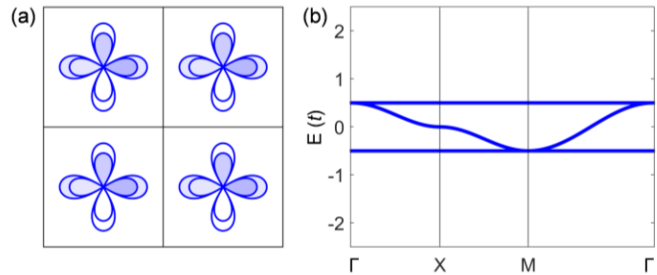

FIG. S4. (a) TFBs in a square lattice with ($d_{x^2-y^2}$, $p_x$, $p_y$) orbitals. (b) Band structure of (a) having the NN hopping integrals $t_{pp\sigma} = \frac{1}{4}, t_{dp\sigma} = \frac{1}{2\sqrt{6}}, t_{dd\sigma} = -\frac{1}{6}$.

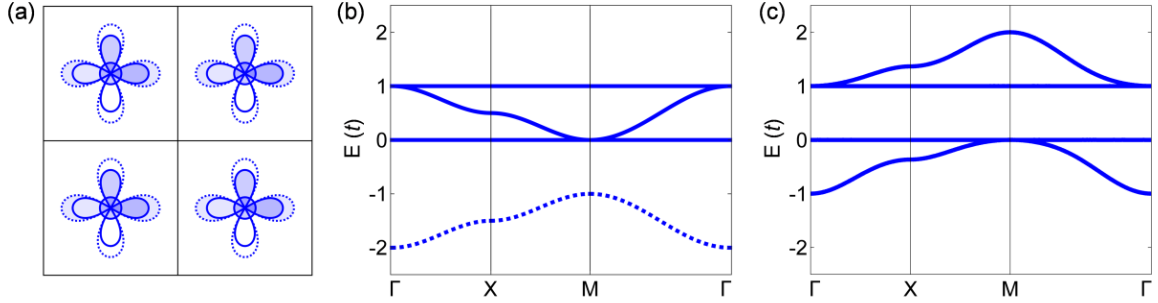

FIG. S5. (a) Dual FBs in a square lattice with ($s$, $p_x$, $p_y$, $d_{x^2-y^2}$) orbitals. (b) Band structure of (a) has NN hopping integrals $t_{ss\sigma} = -\frac{1}{8}, t_{pp\sigma} = \frac{1}{4}, t_{sp\sigma} = \frac{1}{4\sqrt{2}}, t_{dd\sigma} = -\frac{1}{6}$, and onsite energy $\varepsilon_s = \varepsilon_p = \frac{1}{2}$, $\varepsilon_d = -\frac{3}{2}$, which exhibits three upper bands (solid lines) from $s$, $p_x$, $p_y$ orbitals, and one lower band (dashed line) from $d_{x^2-y^2}$ orbital. (c) Yin-Yang FBs in a checkerboard lattice designed from hopping integrals $t_{dd\sigma} = -\frac{1}{2}, t_{sd\sigma} = \frac{1}{4}, t_{pd\sigma} = \frac{\sqrt{2}}{4}$, and onsite energy $\varepsilon_d = \frac{1}{2}$.

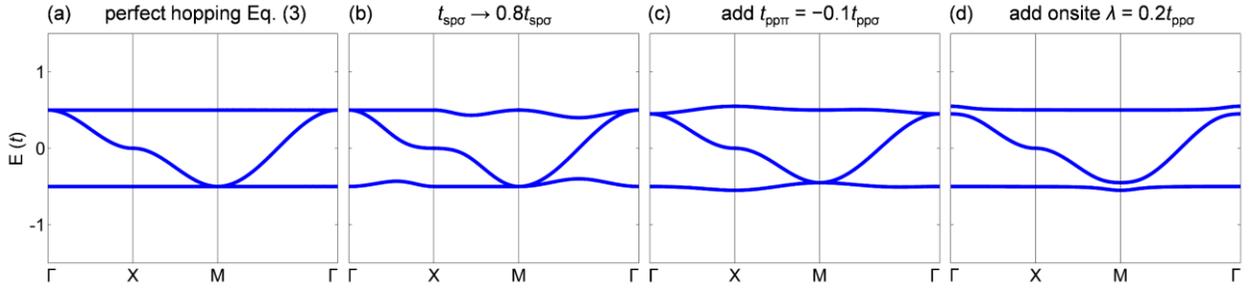

FIG. S6. Illustration of finite dispersion of TFBs in a $sp^2$ square lattice developed with varying hopping parameters. (a) Bands with perfect hopping parameters in Eq. (3), (b) Bands with reduced $0.8t_{sp\sigma}$, (c) Bands with enhanced $t_{pp\pi} = -0.1t_{pp\sigma}$, and (d) Bands with onsite SOC $\lambda = 0.2t_{pp\sigma}$.

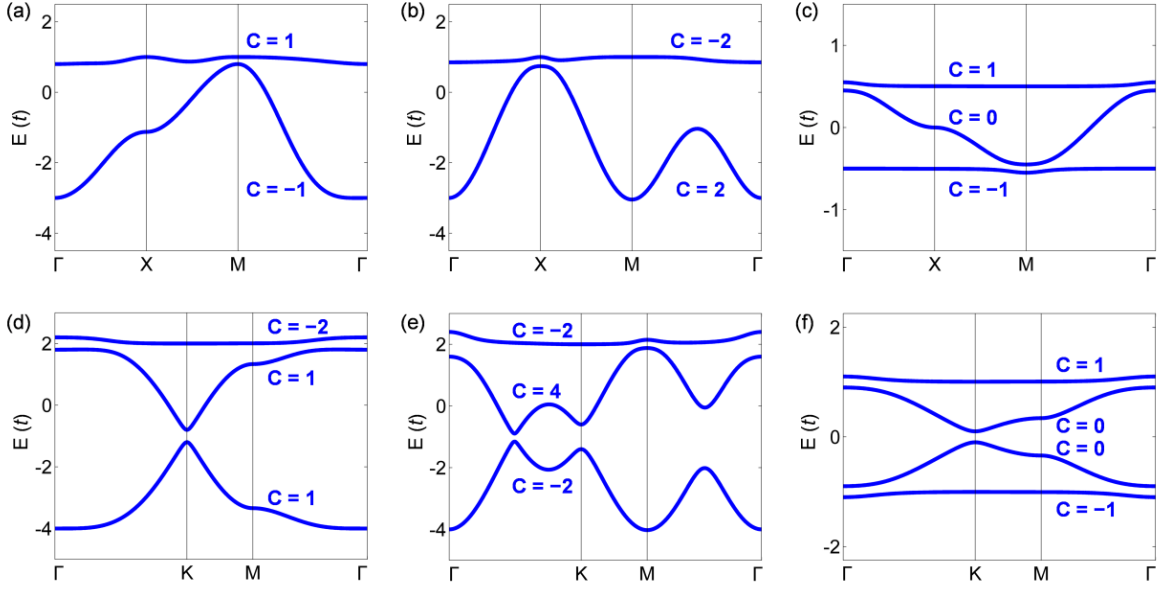

FIG. S7. Band structures and Chern numbers with onsite spin-orbit coupling (SOC, $\lambda$). The Chern number is calculated by integration of Berry curvature in the FBZ in one spin channel. The Chern numbers in the other spin channel are of opposite sign. (a) $sp$-square lattice with $\lambda = 0.8$, (b) $sd$-square lattice with $\lambda = 0.3$, (c) $sp^2$-square lattice with $\lambda = 0.05$, (d) $sp^2$-trigonal lattice with $\lambda = 0.2$, (e) $sd^2$-trigonal lattice with $\lambda = 0.2$, (f) $d^2$-hexagonal lattice with $\lambda = 0.05$. The onsite-SOC matrix elements are given in Table S1.

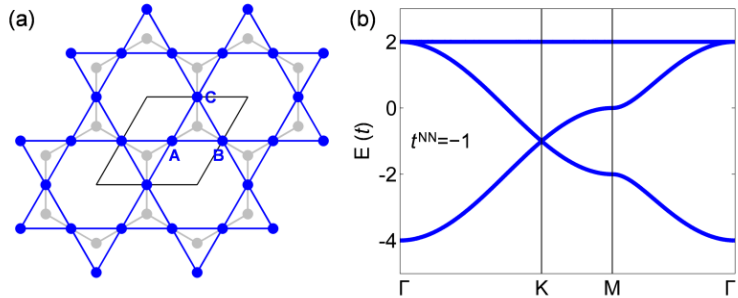

FIG. S8. (a) Illustration of Kagome lattice (blue lines and dots), with three sites (A, B and C) in the unit cell (black thin lines), as LG of hexagonal lattice (gray lines and dots). (b) Band structure of Kagome lattice with the NN hopping integral $t^{NN} = -1$.

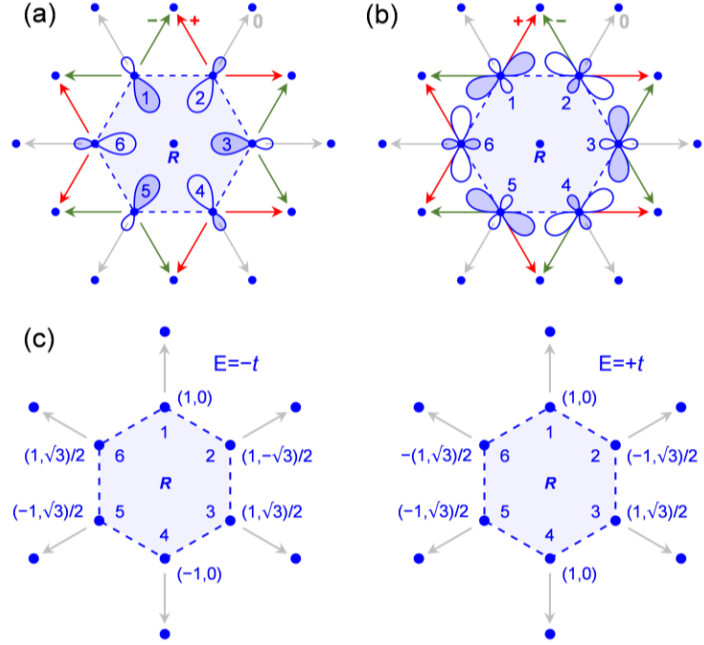

FIG. S9. The CLS on a plaquette for orbital-designed TFBs, illustrating overall zero outward hoppings. Red, green, and gray arrows represent positive, negative, and zero hopping integrals, respectively. (a) $sp^2$-trigonal lattice in Fig. 5(a)-(c), (b) $sd^2$-trigonal lattice in Fig. 5(d)-(f), (c) $d^2$-hexagonal lattice in Fig. 6. See detailed discussions in the Section II of Supplemental Text.

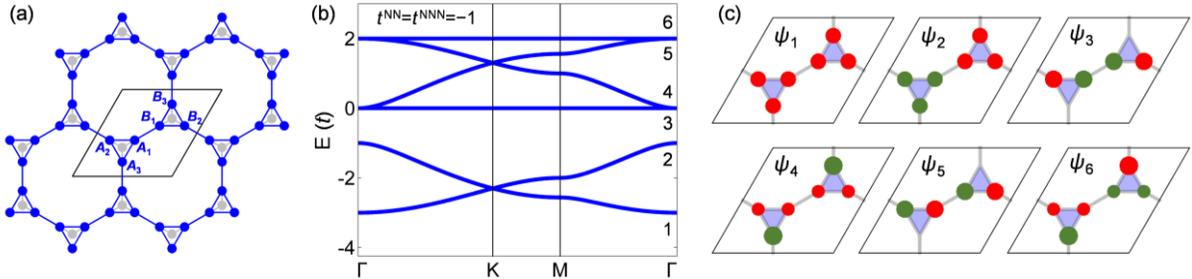

FIG. S10. (a) Illustration of diatomic-Kagome lattice (blue lines and dots), with sites ($A_1$, $A_2$, $A_3$, $B_1$, $B_2$, and $B_3$) in the unit cell (black lines) as a generalized LG of hexagonal lattice (gray lines and dots), consisting of two copies of LG. (b) Band structure of (a) with equal NN and NNN hopping integral $t^{NN} = t^{NNN} = -1$. (c) Illustration of Γ-point lattice wavefunctions for the six states in (b). Red and green color represent the positive and negative nodes of the lattice wavefunctions on six sites, respectively, to show equivalent orbital symmetry.

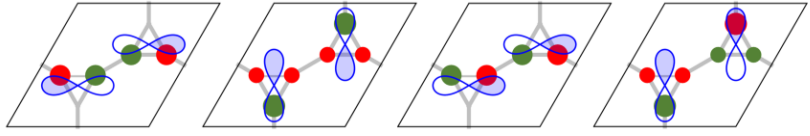

FIG. S11. Illustration of the ($p_x$, $p_y$) orbital basis in a hexagonal lattice having the same symmetry of four Γ-point lattice wavefunctions of a diatomic-Kagome lattice in Fig. S10(c).

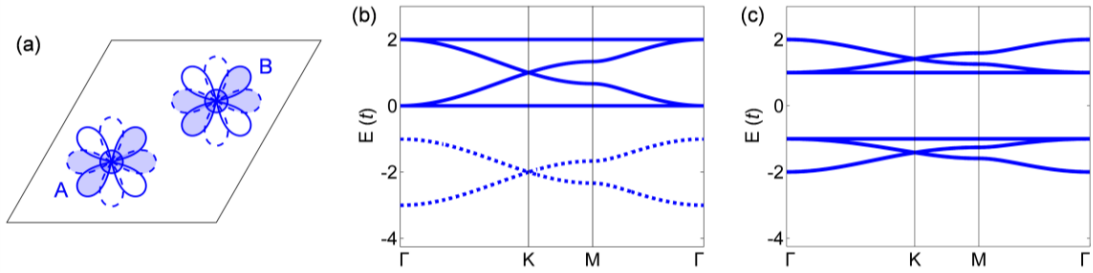

FIG. S12. (a) Hexagonal lattice with atomic orbitals ($s$, $d_{xy}$, $d_{x^2-y^2}$) on each lattice site. (b) Band structure of (a) calculated with $t_{ss\sigma} = -\frac{1}{3}$, $t_{dd\sigma} = -\frac{8}{9}$, and onsite energy $\varepsilon_s = -2$, $\varepsilon_d = 1$. The lower Dirac bands (dashed lines) have only the $s$-component. (c) The Yin-Yang Kagome bands calculated with $t_{ss\sigma} = -\frac{2}{3}$, $t_{dd\sigma} = -\frac{8}{9}$, and $t_{sd\sigma} = -\frac{4}{3\sqrt{3}}$.

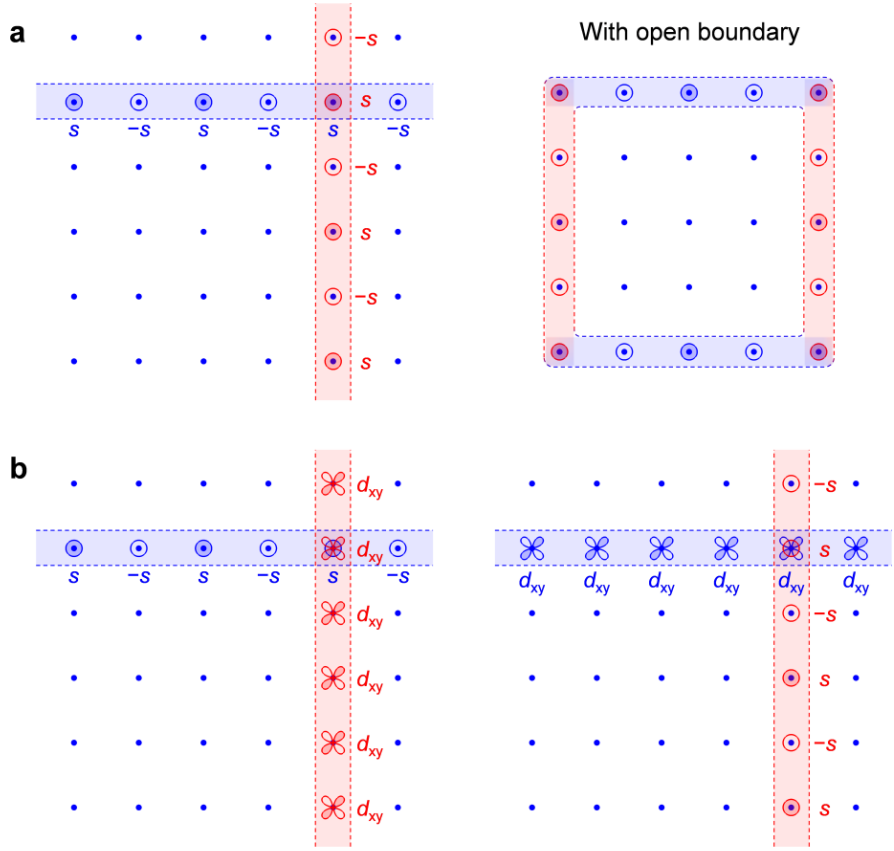

FIG. S13. Illustration of the NLS for TFBs in Fig. 2. (a) corresponds to FB touching point at M in Fig. 2(c). Left panel: NLS under periodic boundary condition. Right panel: NLS under open boundary condition. (b) corresponds to FB touching points at X($\pi$, 0) (left panel) and X(0, $\pi$) (right panel) in Fig. 2(g). Blue dots represent lattice sites of square lattice. See details in Section III of Supplemental Text.

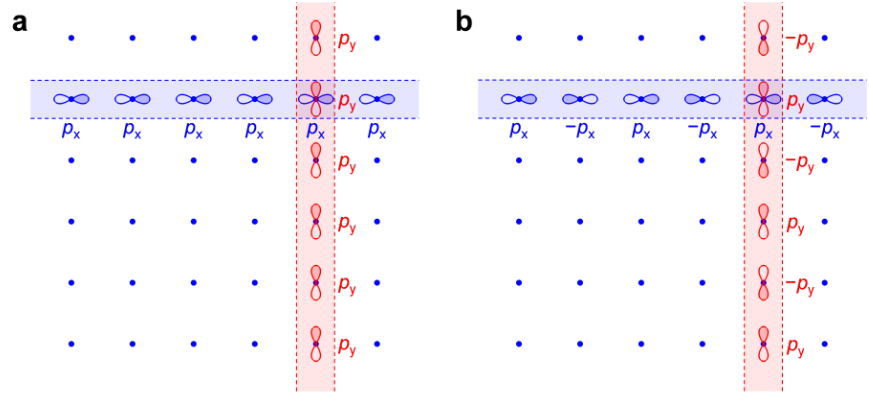

FIG. S14. Illustration of the NLS for TFBs in Fig. 3: (a) corresponds to upper FB touching point at Γ in Fig. 3(c), (b) corresponds to lower FB touching point at M in Fig. 3(c). Blue dots represent lattice sites of square lattice. See details in Section III of Supplemental Text.

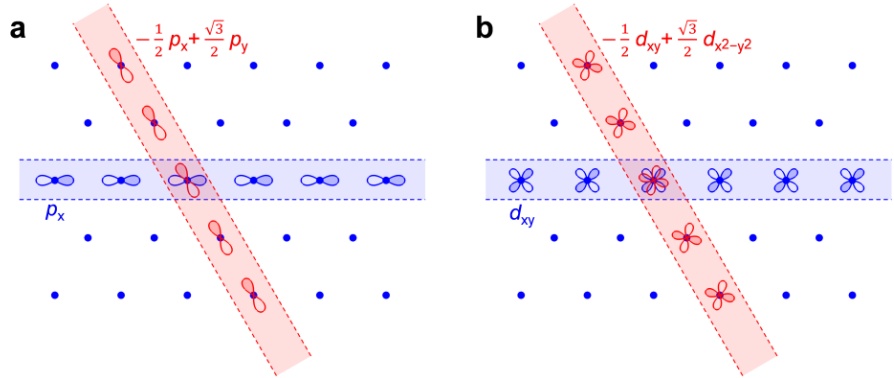

FIG. S15. Illustration of the NLS for TFBs in Fig. 5: (a) corresponds to FB touching point at Γ in Fig. 5(c), (b) corresponds to FB touching points at Γ (or M) in Fig. 5(f). Blue dots represent lattice sites of trigonal lattice. See details in Section III of Supplemental Text.

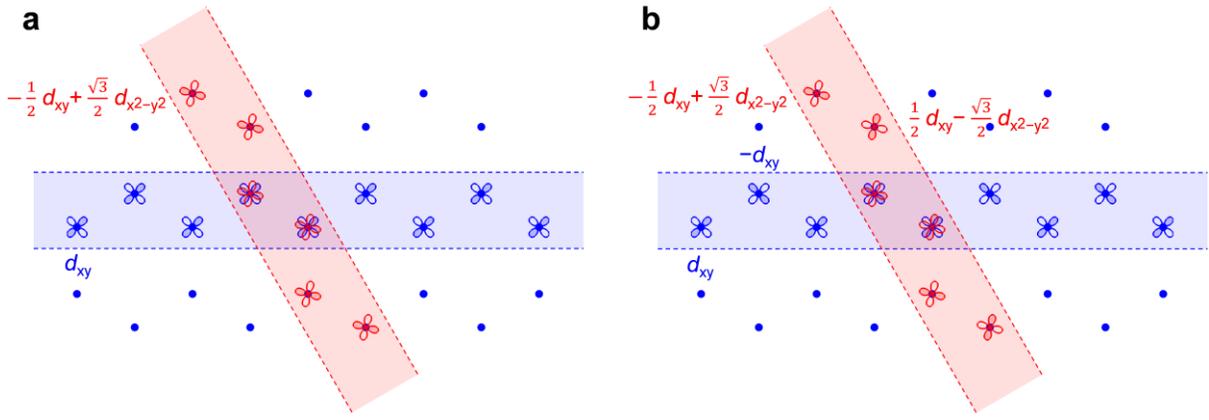

FIG. S16. Illustration of the NLS for TFBs in Fig. 6: (a) corresponds to upper FB touching point at $\Gamma$ in Fig. 6(c), (b) corresponds to lower FB touching point at $\Gamma$ in Fig. 6(c). Blue dots represent lattice sites of hexagonal lattice. See details in Section III of Supplemental Text.

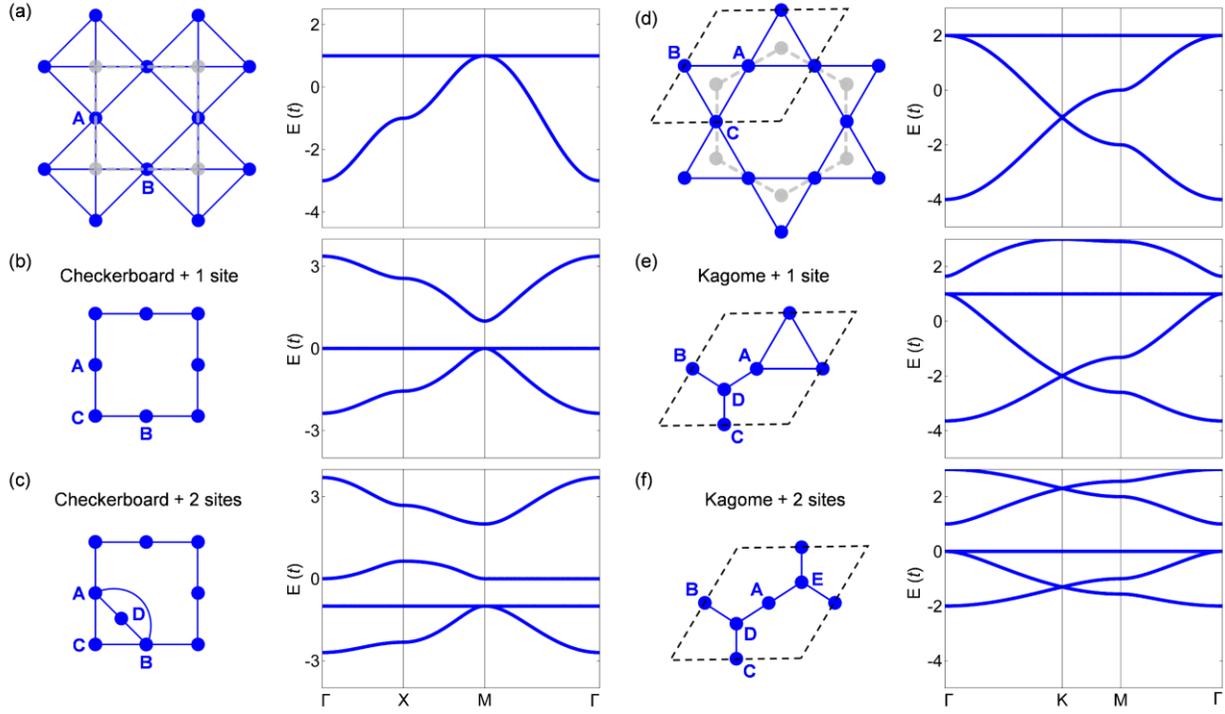

FIG. S17. TFBs in some general/decorated/partial/subdivision LG lattices. (a-c) General LG lattices based on adding extra sites in a checkerboard lattice with NN hopping integral $t$. (b) The onsite energy on C is nonzero: $\varepsilon_C = t$. (d-f) General LG lattices based on adding extra sites in a Kagome lattice with NN hopping integral $t$. (e) The NN hopping integral to D is $t' = \sqrt{2}t$. (f) The onsite energy on D and E are nonzero: $\varepsilon_{D,E} = 0.5t$.

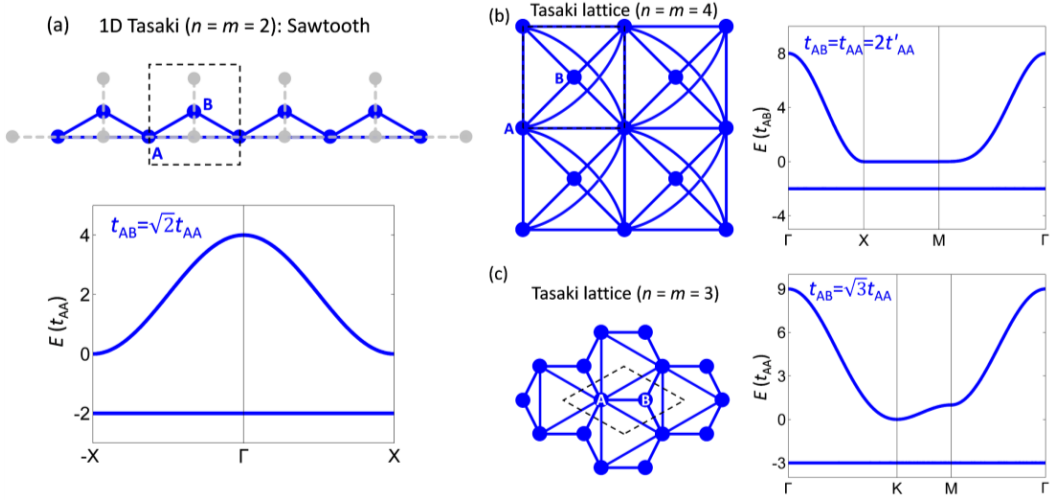

FIG. S18. Topologically fragile/trivial isolated FBs in Tasaki lattices. (a) Top panel: 1D sawtooth lattice viewed as a LG (blue dots and lines) of a side-chain (grey dashed lines and dots) following the standard LG construction. It was originally constructed by cell construction developed by Tasaki with cell dimension $n = m = 2$ (dashed rectangle): a cell contains $n = 2$ external sites A and one internal site B, and $m = 2$ cells share the same external site A. Lower panel: Band structure, based on hopping integral $t_{AB} = \sqrt{2} t_{AA}$, showing an isolated FB. (b) 2D Tasaki lattice ($n = m = 4$) and band structure with hopping integral $t_{AB} = t_{AA} = 2t'_{AA}$, which is related to a LG, i.e., checkerboard lattice [27]. (c) 2D Tasaki lattice ($n = m = 3$) and band structure with hopping integral $t_{AB} = \sqrt{3} t_{AA}$.

Table S1. Matrix elements of onsite SOC in the basis of orbitals.

| Orbital | $p_x$ | $p_y$ | $p_z$ |
|---------|-------|-------|-------|
| $p_x$   | 0     | $-i\lambda$ | $i\lambda$ |
| $p_y$   | $i\lambda$ | 0 | $-i\lambda$ |
| $p_z$   | $-i\lambda$ | $i\lambda$ | 0 |

| Orbital | $d_{xy}$ | $d_{x^2-y^2}$ | $d_{xz}$ |
|---------|----------|---------------|----------|
| $d_{xy}$ | 0 | $2i\lambda$ | $-i\lambda$ |
| $d_{x^2-y^2}$ | $-2i\lambda$ | 0 | $i\lambda$ |
| $d_{xz}$ | $i\lambda$ | $-i\lambda$ | 0 |

# Supplemental Text

**I. Lattice model for FB construction.**

To date, FB has been mostly constructed by lattice models, with one *s*-orbital per lattice site, without explicit consideration of orbital symmetry. Among them, FB lattices constructed from line graph (LG) [1-20], cell construction [21-26], and compact localized states (CLS) [24,27,28] are generic models, which have been independently developed but related. In the following, we briefly review the LG and cell models and discuss their relationship.

**1. Mielke lattice based on line graph.**

A standard LG is made by connecting the centers of edges that share a common vertex of the original graph [1-20]. The LG of bipartite lattices has been demonstrated to have the topological FB (TFB), characterized by a singular touching point with a dispersive band [19,28,29]. For example, as shown in Fig. S17(a), the LG of square lattice is checkerboard lattice [4], exhibiting a FB touching a dispersive band at M. As shown in Fig. S17(d), the LG of hexagonal lattice is Kagome lattice [1-3], exhibiting a FB touching Dirac bands at Γ; the breathing-Kagome [7-10], diatomic-Kagome [11,12], coloring-triangle [13], and the diamond-octagon lattices [14,15] are among other studied LG lattices with TFBs.

The lattices containing a LG sublattice of bipartite lattices also host TFBs, where these lattices are called general/partial/decorated/subdivision LG lattices in literatures [16-20]. For example, if a new site is added to checkerboard lattice [Fig. S17(b)], such as a Lieb lattice, it hosts also a TFB. If another new site is further added to Lieb lattice, the TFB persists [Fig. S17(c)]. Similarly, adding extra sites to Kagome lattice can keep the original Kagome bands intact, as shown in Figs. S17(d-f). More specific examples can be found in various literatures [13,16-20]. This indicates that the FBs in these seemingly "non-LG" lattices can be considered still rooted from LG construction. On the other hand, FBs in LG lattices of non-bipartite lattices are isolated from other bands, which are topologically trivial/fragile [5,14,19].

## 2. Tasaki lattice based on cell construction.

Cell construction is made as follows: each cell contains $n$ external sites and one internal site, then the $m$ cells sharing the same external site are connected, forming ($n$, $m$) Tasaki lattice with topologically trivial/fragile FBs [21-26]. For example, as shown in Fig. S18(a), the 1D Tasaki lattice with $n = m = 2$ (i.e., sawtooth lattice) exhibits a trivial FB [24,29], as it is isolated from the dispersive band; as shown in Fig. S18(b,c), the FBs in 2D Tasaki lattices do not have singular band touching/crossing points [25], indicating they are also topologically trivial/fragile. It is interesting to note that the Tasaki lattices are either directly LG lattices or related to LG lattices: the 1D Tasaki saw-tooth lattice [blue dots and lines in Fig. S18(a)] is a LG of a 1D side-chain lattice [grey dots and lines in Fig. S18(a)]; similarly, 2D Tasaki lattices are also related to LGs [27].

## II. Compact localized state of orbital-designed FBs.

The CLS is an important physical property of FB, as a direct manifestation of destructive interference (phase cancellation) of lattice wavefunctions in real space in association with the FB. Conversely, it offer another generic lattice model to construct FB [24,27,28], as mentioned above. In lattice models without considering orbital symmetry, the CLS is resulted from destructive lattice hopping induced solely by lattice symmetry, as reflected by the alternating nodal signs of Bloch wavefunction on an even-edged plaquette. The CLS of orbital-designed FBs in non-LG lattices, as we developed here, is more complex, because the phase cancellation of Bloch wavefunctions involves both lattice and orbital symmetries, as we illustrate in detail below.

### 1. Square lattice.

*sp model*. The Bloch state of TFB in Fig. 2(c) is calculated as $\psi_k^{FB} = i\frac{1}{\sqrt{2}}\sin\frac{k_1+k_2}{2}|s\rangle + \cos\frac{k_1}{2}\cos\frac{k_2}{2}|p\rangle$ with $k_n = \mathbf{k} \cdot \mathbf{a}_n$ ($\mathbf{a}_n$, lattice vector; $n = 1, 2$), whose Fourier transformation $\psi_R^{FB} = \int_{BZ} d\mathbf{k} e^{-i\mathbf{k}\cdot\mathbf{R}}\psi_k^{FB}$ produces a real-space CLS on a square plaquette centered at $\mathbf{R}$ [Fig. 2(d)]. It consists of nodal wavefunctions of $\frac{|p\rangle}{4}$, $-\frac{|s\rangle}{2\sqrt{2}} + \frac{|p\rangle}{4}$, $\frac{|p\rangle}{4}$, and $\frac{|s\rangle}{2\sqrt{2}} + \frac{|p\rangle}{4}$ at four vertices of the

plaquette, respectively. Electron hoppings outward from the CLS to its surrounding lattice sites are completely forbidden, which can be shown by analyzing hoppings based on Eq. (2). As shown in Fig. 2(d), electron hoppings outward from the CLS of *sp*-square lattice include that from four (eight) sites to its NNN (NN) sites, based on hopping integrals in Eq. (2). First, each of the four NNN sites has the electron hopping coming from one site inside the CLS: (i) the hopping from the CLS to its top-left site is from site 1, which is $\frac{1}{4}t_{pp\pi}^{NNN}=0$ since $t_{pp\pi}^{NNN}$ is zero (likewise, from site 3 to its bottom-right site); (ii) the hopping from the CLS to its top-right site is from site 2, which is $-\frac{1}{2\sqrt{2}}\left(t_{ss\sigma}^{NNN}+t_{sp\sigma}^{NNN}\right)+\frac{1}{4}\left(-t_{sp\sigma}^{NNN}+t_{pp\sigma}^{NNN}\right)=0$ due to the cancellation of four nonzero terms (likewise, from site 4 to its bottom-left site). Second, each of the eight NN sites has electron hopping coming from two sites inside the CLS: the hoppings to the site right above site 1 are from site 1 and 2, which is $\frac{1}{4}\left[\frac{1}{2}\left(t_{pp\sigma}^{NN}+t_{pp\pi}^{NN}\right)-\frac{1}{\sqrt{2}}t_{sp\sigma}^{NN}\right]=-\frac{t}{8\sqrt{2}}$ and $-\frac{1}{2\sqrt{2}}t_{ss\sigma}^{NNN}+\frac{1}{4}t_{pp\pi}^{NNN}=\frac{t}{8\sqrt{2}}$, respectively, which cancel out with each other.

*sd model*. The TFB in *sd*-orbital model supports a CLS on a square plaquette with linearly combined *s* and *d* orbitals on its vertices [Fig. 2(h)], whose outward hoppings also vanish. The FB Bloch wavefunction in *sd*-square lattice is $\psi_k^{FB}=\sin\frac{k_x}{2}\sin\frac{k_y}{2}|s\rangle+\cos\frac{k_x}{2}\cos\frac{k_y}{2}|d_{xy}\rangle$. Its CLS consists of $\frac{1}{4}\left(|s\rangle+|d_{xy}\rangle\right)$ and $\frac{1}{4}\left(-|s\rangle+|d_{xy}\rangle\right)$ nodes, respectively, on the corner (1, 3) and (2, 4) of a square plaquette, where all outward electron hoppings vanish, as illustrated in Fig. 2(h). For example, the hopping from the CLS to its top-right NNN site is from site 2, which is $\frac{1}{4}\left(-t_{ss\sigma}^{NNN}+\frac{3}{4}t_{dd\sigma}^{NNN}\right)=0$; the hopping to the site right above site 2 is from site 1 and 2, which are $\frac{1}{4}\left(t_{ss\sigma}^{NNN}+\frac{\sqrt{3}}{2}t_{sd\sigma}^{NNN}\right)+\frac{1}{4}\left(\frac{3}{4}t_{dd\sigma}^{NNN}+\frac{\sqrt{3}}{2}t_{sd\sigma}^{NNN}\right)=-\frac{t}{4}$ and $-\frac{1}{4}t_{ss\sigma}^{NN}+\frac{1}{4}t_{dd\pi}^{NN}=\frac{t}{4}$, respectively, opposite of each other.

*sp² model.* As illustrated in Fig. 3(d), the lower FB supports a CLS on a square plaquette with bonding nodal wavefunctions $|s\rangle + \frac{1}{\sqrt{2}}(|p_x\rangle - |p_y\rangle)$, $|s\rangle - \frac{1}{\sqrt{2}}(|p_x\rangle + |p_y\rangle)$, $|s\rangle + \frac{1}{\sqrt{2}}(-|p_x\rangle + |p_y\rangle)$, and $|s\rangle + \frac{1}{\sqrt{2}}(|p_x\rangle + |p_y\rangle)$ at four vertices labled as 1, 2, 3, and 4 (left panel), respectively; while the upper-FB CLS consists of four anti-bonding vertex states $|s\rangle + \frac{1}{\sqrt{2}}(|p_x\rangle - |p_y\rangle)$, $-|s\rangle + \frac{1}{\sqrt{2}}(|p_x\rangle + |p_y\rangle)$, $|s\rangle + \frac{1}{\sqrt{2}}(-|p_x\rangle + |p_y\rangle)$, and $-|s\rangle - \frac{1}{\sqrt{2}}(|p_x\rangle + |p_y\rangle)$ on vertex 1, 2, 3, and 4, respectively (right panel). The vanishing outward electron hoppings from the CLS of *sp²*-square lattice include eight hoppings to its NN sites. The nodal wavefunction on site $m$ ($m$ = 1, 2, 3, 4) can be written as $a_m|s\rangle + b_m|p_x\rangle + c_m|p_y\rangle$, whose upward, downward, leftward, and rightward hoppings are $a_m(t_{ss\sigma} + t_{sp\sigma}) + c_m(-t_{sp\sigma} + t_{pp\sigma})$ with $m$ = 1, 2, $a_m(t_{ss\sigma} - t_{sp\sigma}) + c_m(t_{sp\sigma} + t_{pp\sigma})$ with $m$ = 3, 4, $a_m(t_{ss\sigma} - t_{sp\sigma}) + b_m(t_{sp\sigma} + t_{pp\sigma})$ with $m$ = 1, 4, $a_m(t_{ss\sigma} + t_{sp\sigma}) + b_m(-t_{sp\sigma} + t_{pp\sigma})$ with $m$ = 2, 3, respectively, which all equal to zero based on hopping integrals in Eq. (3). For example, the one from site 1 to the site above is $(t_{ss\sigma} + t_{sp\sigma}) - \frac{1}{\sqrt{2}}(-t_{sp\sigma} + t_{pp\sigma}) = 0$. It confirms that the orbital symmetry undelines the destructive interference of Bloch wavefunctions for both FBs.

**2. Trigonal lattice.**

*sp² model.* The CLS of the FB in a *sp²*-trigonal lattice is on a hexagon plaquette [Fig. S9(a)] with $|s\rangle + \frac{1}{\sqrt{6}}|p_x\rangle - \frac{1}{\sqrt{2}}|p_y\rangle$, $-|s\rangle + \frac{1}{\sqrt{6}}|p_x\rangle + \frac{1}{\sqrt{2}}|p_y\rangle$, $|s\rangle - \frac{2}{\sqrt{6}}|p_x\rangle$, $-|s\rangle + \frac{1}{\sqrt{6}}|p_x\rangle - \frac{1}{\sqrt{2}}|p_y\rangle$, $|s\rangle + \frac{1}{\sqrt{6}}|p_x\rangle + \frac{1}{\sqrt{2}}|p_y\rangle$, and $-|s\rangle - \frac{2}{\sqrt{6}}|p_x\rangle$ on vertex 1, 2, 3, 4, 5, and 6, respectively. Electron hoppings from the CLS to its surrounding NN lattice sites cancel out, and those to surrounding NNN lattice sites are zero, as indicated by arrows in Fig. S9(a).

*sd² model*. The CLS of the FB in a *sd²*-trigonal lattice is on a hexagon plaquette [Fig. S9(b)]

with $|s\rangle + \frac{1}{\sqrt{2}}|d_{x^2-y^2}\rangle + \frac{\sqrt{6}}{2}|d_{xy}\rangle$ , $-|s\rangle - \frac{1}{\sqrt{2}}|d_{x^2-y^2}\rangle + \frac{\sqrt{6}}{2}|d_{xy}\rangle$ , $|s\rangle - \sqrt{2}|d_{x^2-y^2}\rangle$ ,

$-|s\rangle - \frac{1}{\sqrt{2}}|d_{x^2-y^2}\rangle - \frac{\sqrt{6}}{2}|d_{xy}\rangle$, $|s\rangle + \frac{1}{\sqrt{2}}|d_{x^2-y^2}\rangle - \frac{\sqrt{6}}{2}|d_{xy}\rangle$, and $-|s\rangle + \sqrt{2}|d_{x^2-y^2}\rangle$ at vertex 1, 2, 3, 4, 5, and 6, respectively. Electron hoppings from the CLS to its surrounding NN lattice sites cancel out, and that to surrounding NNN lattice sites are zero, as indicated by arrows in Fig. S9(b).

### 3. Hexagonal lattice.

As shown in left panel of Fig. S9(c), the CLS of upper FB in $d^2$-hexagonal lattice is on a hexagon plaquette with nodal wavefunctions $|d_{xy}\rangle$, $\frac{1}{2}(|d_{xy}\rangle - \sqrt{3}|d_{x^2-y^2}\rangle)$, $\frac{1}{2}(|d_{xy}\rangle + \sqrt{3}|d_{x^2-y^2}\rangle)$, $-|d_{xy}\rangle$, $\frac{1}{2}(-|d_{xy}\rangle + \sqrt{3}|d_{x^2-y^2}\rangle)$, and $\frac{1}{2}(|d_{xy}\rangle + \sqrt{3}|d_{x^2-y^2}\rangle)$ on vertex 1, 2, 3, 4, 5, and 6, respectively. Hoppings from the CLS to all surrounding lattice sites are zero.

As shown in right panel of Fig. S9(c), the CLS of lower FB in in $d^2$-hexagonal lattice is on a hexagon plaquette with nodal wavefunctions $|d_{xy}\rangle$, $-\frac{1}{2}(|d_{xy}\rangle - \sqrt{3}|d_{x^2-y^2}\rangle)$, $\frac{1}{2}(|d_{xy}\rangle + \sqrt{3}|d_{x^2-y^2}\rangle)$, $|d_{xy}\rangle$, $\frac{1}{2}(-|d_{xy}\rangle + \sqrt{3}|d_{x^2-y^2}\rangle)$, and $-\frac{1}{2}(|d_{xy}\rangle + \sqrt{3}|d_{x^2-y^2}\rangle)$ on vertex 1, 2, 3, 4, 5, and 6, respectively. Hoppings from the CLS to all surrounding lattice sites are zero.

### III. Noncontractible loop state (NLS) of orbital-designed TFBs.

When a FB singularly touches a dispersive band at $k_0 = (k_{x0}, k_{y0})$ point, one can perform Fourier transformation of FB Bloch wave functions $\psi_k^{FB}$ at all points along a $K$ path passing through the $k_0$ point, to derive the real-space wave function $\psi_r^{FB} = \int_{K_{path}} dk e^{ik \cdot r} \psi_k^{FB}$. This real-space FB state is extended along the direction reciprocal to $K$ path, and localized along the orthogonal direction. It is known as NLS [28,29], exhibiting a robust edge state with open boundary condition. The emergence of NLSs demonstrates the topology/singularity of orbital-designed FBs. The calculated NLSs $\psi_{NLS} = \psi_r^{FB}$ for all the orbital-FB models presented in this work are illustrated in Figs. S13-S16, as explained in detail below.

### 1. Square lattice.

*sp model.* The FB of the *sp* square lattice model [Fig. 2(a)-(d)] touches a dispersive band at M point. As shown in the left panel of Fig. S13(a), two NLSs of $\psi_{NLS} = \sum_{N_x, N_y=0} (-1)^{|N_x|} |s\rangle$ and $\psi_{NLS} = \sum_{N_x=0, N_y} (-1)^{|N_y|} |s\rangle$ are constructed by the Fourier integrals along $k_y$ and $k_x$ path containing M point, respectively. The position of lattice sites is $(N_x, N_y)a$ with lattice constant $a$. The localization of NLSs can be further checked by the destructive interference of electron hopping based on Eq. (2). For a finite system, the NLS corresponds to a robust boundary state, as shown in the right panel of Fig. S13(a).

*sd model.* The FB of *sd* square lattice model [Fig. 2(e)-(h)] touches a dispersive band at two X points of $(\pi, 0)$ and $(0, \pi)$. As shown in the left panel of Fig. S13(b), two NLSs of $\psi_{NLS} = \sum_{N_x, N_y=0} (-1)^{|N_x|} |s\rangle$ and $\psi_{NLS} = \sum_{N_x=0, N_y} |d_{xy}\rangle$ are constructed by the Fourier integrals along $k_y$ and $k_x$ path containing $X(\pi, 0)$ point, respectively. As shown in the right panel of Fig. S13(b), two NLSs of $\psi_{NLS} = \sum_{N_x, N_y=0} |d_{xy}\rangle$ and $\psi_{NLS} = \sum_{N_x=0, N_y} (-1)^{|N_x|} |s\rangle$ are constructed by the Fourier integrals along $k_y$ and $k_x$ path containing $X(0, \pi)$ point, respectively.

*sp² model.* The upper and lower FB of *sp²* square lattice model (Fig. 3) touches a dispersive band at Γ and M point, respectively. As shown in Fig. S14(a), two upper-FB NLSs of $\psi_{NLS} = \sum_{N_x, N_y=0} |p_x\rangle$ and $\psi_{NLS} = \sum_{N_x=0, N_y} |p_y\rangle$ are constructed by the Fourier integrals along $k_y$ and $k_x$ path containing Γ point, respectively. As shown in Fig. S14(b), two lower-FB NLSs of $\psi_{NLS} = \sum_{N_x, N_y=0} (-1)^{|N_x|} |p_x\rangle$ and $\psi_{NLS} = \sum_{N_x=0, N_y} (-1)^{|N_y|} |p_y\rangle$ are constructed by the Fourier integrals along $k_y$ and $k_x$ path containing M point, respectively.

## 2. Trigonal lattice.

*sp² model.* The FB of *sp²* trigonal lattice model [Fig. 5(a)-(c)] touches a dispersive band at Γ point. As shown in Fig. S15(a), the NLS with orbital $|p_x\rangle$ (orbital $-\frac{1}{2}|p_x\rangle + \frac{\sqrt{3}}{2}|p_y\rangle$) on every NN lattice sites along $x$ direction ($-\frac{1}{2}x + \frac{\sqrt{3}}{2}y$ direction) is constructed by the Fourier integral along $k_y$ path ($\frac{\sqrt{3}}{2}k_x + \frac{1}{2}k_y$ path) containing Γ point.

*sd² model.* The FB of *sd²* trigonal lattice model [Fig. 5(d)-(f)] touches a dispersive band at Γ and M points. As shown in Fig. S15(b), the NLS with orbital $|d_{xy}\rangle$ (orbital $-\frac{1}{2}|d_{xy}\rangle + \frac{\sqrt{3}}{2}|d_{x^2-y^2}\rangle$) on NN lattice sites along $x$ direction ($-\frac{1}{2}x + \frac{\sqrt{3}}{2}y$ direction) is constructed by the Fourier integral along $k_y$ path ($\frac{\sqrt{3}}{2}k_x + \frac{1}{2}k_y$ path) containing Γ and M points.

## 3. Hexagonal lattice.

*d² model.* The upper and lower FB of *d²* square lattice model (Fig. 6) touches a dispersive band at Γ point. As shown in Fig. S16(a), the upper-FB NLS with orbital $|d_{xy}\rangle$ (orbital $-\frac{1}{2}|d_{xy}\rangle + \frac{\sqrt{3}}{2}|d_{x^2-y^2}\rangle$) on zigzag NN lattice sites along $x$ direction ($-\frac{1}{2}x + \frac{\sqrt{3}}{2}y$ direction) is constructed by the Fourier integral along $k_y$ path ($\frac{\sqrt{3}}{2}k_x + \frac{1}{2}k_y$ path) containing Γ point. As shown in Fig. S16(b), the lower-FB NLS with orbitals $|d_{xy}\rangle$ and $-|d_{xy}\rangle$ (orbitals $-\frac{1}{2}|d_{xy}\rangle + \frac{\sqrt{3}}{2}|d_{x^2-y^2}\rangle$ and $\frac{1}{2}|d_{xy}\rangle -$

$\frac{\sqrt{3}}{2}|d_{x^2-y^2}\rangle$) alternating on zigzag NN lattice sites along $x$ direction ($-\frac{1}{2}x + \frac{\sqrt{3}}{2}y$ direction) is constructed by the Fourier integral along $k_y$ path ($\frac{\sqrt{3}}{2}k_x + \frac{1}{2}k_y$ path) containing $\Gamma$ point.

## IV. Momentum-space Hamiltonians without SOC.

Note: Please refer to Computational Methods in main text for Hamiltonians with SOC.

**1.1 $sp$-square lattice**: lattice vectors $\boldsymbol{a}_1 = (a,0)$, $\boldsymbol{a}_2 = (0,a)$ ; $k_1 = \boldsymbol{k}\cdot\boldsymbol{a}_1$, $k_2 = \boldsymbol{k}\cdot\boldsymbol{a}_2$; basis $\{s, p\}$.

$$H = \begin{pmatrix} 2t_{ss\sigma}^{NN}(\cos k_1 + \cos k_2) + 4t_{ss\sigma}^{NNN} \cos k_1 \cos k_2 & i\sqrt{2} t_{sp\sigma}^{NN}(\sin k_1 + \sin k_2) + i2 t_{sp\sigma}^{NNN} \sin(k_1 + k_2) \\ \dagger & (t_{pp\sigma}^{NN} + t_{pp\pi}^{NN})(\cos k_1 + \cos k_2) + 2t_{pp\sigma}^{NNN} \cos(k_1 + k_2) + 2t_{pp\pi}^{NNN} \cos(k_1 - k_2) \end{pmatrix}$$

$\downarrow$

$$H_{FB} = \begin{pmatrix} -(\cos k_1 + \cos k_2 + \cos k_1 \cos k_2) & \dfrac{i}{\sqrt{2}}(\sin k_1 + \sin k_2 + \sin(k_1 + k_2)) \\ \dagger & \cos(k_1 + k_2) \end{pmatrix} \qquad \begin{array}{l} \Gamma : (k_1, k_2) = (0,0) \\ \psi_{E=-3t} = |s\rangle, \psi_{E=t} = |p\rangle \end{array}$$

**1.2 $sd$-square lattice**: basis $\{s, d_{xy}\}$.

$$H = \begin{pmatrix} 2t_{ss\sigma}^{NN}(\cos k_1 + \cos k_2) + 4t_{ss\sigma}^{NNN} \cos k_1 \cos k_2 & -2\sqrt{3} t_{sd\sigma}^{NNN} \sin k_1 \sin k_2 \\ \dagger & 2t_{dd\pi}^{NN}(\cos k_1 + \cos k_2) + (3t_{dd\sigma}^{NNN} + t_{dd\delta}^{NNN})\cos k_1 \cos k_2 \end{pmatrix}$$

$\downarrow$

$$H_{FB} = \begin{pmatrix} -(\cos k_1 + \cos k_2 + \cos k_1 \cos k_2) & \sin k_1 \sin k_2 \\ \dagger & \cos k_1 + \cos k_2 - \cos k_1 \cos k_2 \end{pmatrix} \qquad \Gamma : \psi_{E=-3t} = |s\rangle, \psi_{E=t} = |d\rangle$$

**1.3 Checkerboard lattice**: basis $\{s_A, s_B\}$.

$$H = 2\begin{pmatrix} t^{NNN}\cos k_2 & 2t^{NN}\cos\dfrac{k_1}{2}\cos\dfrac{k_2}{2} \\ \dagger & t^{NNN}\cos k_1 \end{pmatrix} \rightarrow H_{FB} = -\begin{pmatrix} \cos k_2 & 2\cos\dfrac{k_1}{2}\cos\dfrac{k_2}{2} \\ \dagger & \cos k_1 \end{pmatrix} \qquad \Gamma : \psi_{E=-3t} = \dfrac{1}{\sqrt{2}}(|A\rangle + |B\rangle), \psi_{E=t} = \dfrac{1}{\sqrt{2}}(|A\rangle - |B\rangle)$$

**2.1 $sp^2$-square lattice**: basis $\{s, p_x, p_y\}$.

$$H = 2\begin{pmatrix} t_{ss\sigma}(\cos k_1 + \cos k_2) & it_{sp\sigma}\sin k_1 & it_{sp\sigma}\sin k_2 \\ & t_{pp\sigma}\cos k_1 + t_{pp\pi}\cos k_2 & 0 \\ \dagger & & t_{pp\pi}\cos k_1 + t_{pp\sigma}\cos k_2 \end{pmatrix}$$

↓

$$H_{FB} = \frac{1}{4}\begin{pmatrix} -(\cos k_1 + \cos k_2) & i\sqrt{2}\sin k_1 & i\sqrt{2}\sin k_2 \\ & 2\cos k_1 & 0 \\ \dagger & & 2\cos k_2 \end{pmatrix} \qquad \Gamma: \psi_{E=-\frac{t}{2}} = |s\rangle, \psi_{E=\frac{t}{2}} = |p_x\rangle, \psi_{E=\frac{t}{2}} = |p_y\rangle$$

**2.2 $dp^2$-square lattice**: basis $\{d_{x^2-y^2}, p_x, p_y\}$.

$$H = \begin{pmatrix} \frac{1}{2}(3t_{dd\sigma} + t_{dd\delta})(\cos k_1 + \cos k_2) & -i\sqrt{3}t_{pd\sigma}\sin k_1 & i\sqrt{3}t_{pd\sigma}\sin k_2 \\ & 2(t_{pp\sigma}\cos k_1 + t_{pp\pi}\cos k_2) & 0 \\ \dagger & & 2(t_{pp\pi}\cos k_1 + t_{pp\sigma}\cos k_2) \end{pmatrix}$$

↓

$$H_{FB} = \frac{1}{2}\begin{pmatrix} -\frac{1}{2}(\cos k_1 + \cos k_2) & i\frac{1}{\sqrt{2}}\sin k_1 & i\frac{1}{\sqrt{2}}\sin k_2 \\ & \cos k_1 & 0 \\ \dagger & & \cos k_2 \end{pmatrix} \qquad \Gamma: \psi_{E=-\frac{t}{2}} = |d_{x^2-y^2}\rangle, \psi_{E=\frac{t}{2}} = |p_x\rangle, \psi_{E=\frac{t}{2}} = |p_y\rangle$$

**2.3 sp²d-square lattice**: basis $\{p_x, p_y, s, d_{x^2-y^2}\}$.

$$H = \begin{pmatrix} 2(t_{pp\sigma}\cos k_1 + t_{pp\pi}\cos k_2) & 0 & -i2t_{sp\sigma}\sin k_1 & i\sqrt{3}t_{pd\sigma}\sin k_1 \\ & 2(t_{pp\pi}\cos k_1 + t_{pp\sigma}\cos k_2) & -i2t_{sp\sigma}\sin k_2 & -i\sqrt{3}t_{pd\sigma}\sin k_2 \\ & & 2t_{ss\sigma}(\cos k_1 + \cos k_2) & \sqrt{3}t_{sd\sigma}(\cos k_1 - \cos k_2) \\ \dagger & & & \frac{1}{2}(3t_{dd\sigma} + t_{dd\delta})(\cos k_1 + \cos k_2) \end{pmatrix} + \begin{pmatrix} U_p & & & \\ & U_p & & \\ & & U_s & \\ & & & U_d \end{pmatrix}$$

↓

$$H_{FB} = \frac{1}{4}\begin{pmatrix} 2\cos k_1 & 0 & -i\sqrt{2}\sin k_1 & i\sqrt{6}\sin k_1 \\ & 2\cos k_2 & -i\sqrt{2}\sin k_2 & -i\sqrt{6}\sin k_2 \\ & & -(\cos k_1 + \cos k_2) & \sqrt{3}(\cos k_1 - \cos k_2) \\ \dagger & & & -3(\cos k_1 + \cos k_2) \end{pmatrix} + \frac{1}{2}\begin{pmatrix} 1 & & & \\ & 1 & & \\ & & 1 & \\ & & & 1 \end{pmatrix} \qquad \Gamma: \begin{cases} \psi_{E=-t} = |d_{x^2-y^2}\rangle \\ \psi_{E=0} = |s\rangle, \psi_{E=t} = |p_x\rangle, \psi_{E=t} = |p_y\rangle \end{cases}$$

**2.4 Diamond-octagon lattice**: basis $\{s_A, s_B, s_C, s_D\}$.

$$H = \begin{pmatrix} 0 & 2t^{NNN}\cos\frac{k_1}{2} & t^{NN}e^{i\frac{k_2-k_1}{4}} & t^{NN}e^{i\frac{-k_2-k_1}{4}} \\ & 0 & t^{NN}e^{i\frac{k_2+k_1}{4}} & t^{NN}e^{i\frac{-k_2+k_1}{4}} \\ & & 0 & 2t^{NNN}\cos\frac{k_2}{2} \\ & & & 0 \end{pmatrix}$$

$t^{NN} = -t^{NNN} = \frac{1}{2} \to \Gamma: \begin{cases} \psi_{E=-2t} = \frac{1}{2}(|A\rangle + |B\rangle - |C\rangle - |D\rangle), \psi_{E=0} = \frac{1}{2}(|A\rangle + |B\rangle + |C\rangle + |D\rangle) \\ \psi_{E=t} = \frac{1}{\sqrt{2}}(|A\rangle - |B\rangle), \psi_{E=t} = \frac{1}{\sqrt{2}}(|C\rangle - |D\rangle) \end{cases}$

$t^{NN} = t^{NNN} = -\frac{1}{2} \to \Gamma: \begin{cases} \psi_{E=-2t} = \frac{1}{2}(|A\rangle + |B\rangle + |C\rangle + |D\rangle), \psi_{E=0} = \frac{1}{2}(|A\rangle + |B\rangle - |C\rangle - |D\rangle) \\ \psi_{E=t} = \frac{1}{\sqrt{2}}(|A\rangle - |B\rangle), \psi_{E=t} = \frac{1}{\sqrt{2}}(|C\rangle - |D\rangle) \end{cases}$

**3.1 $sp^2$-trigonal lattice**: NN vectors $\mathbf{a}_1 = a(1,0)$, $\mathbf{a}_2 = -\mathbf{a}_1$, $\mathbf{a}_3 = \frac{a}{2}(1,\sqrt{3})$, $\mathbf{a}_4 = -\mathbf{a}_3$, $\mathbf{a}_5 = \mathbf{a}_3 - \mathbf{a}_1$, $\mathbf{a}_6 = -\mathbf{a}_5$; $k_n = \mathbf{k} \cdot \mathbf{a}_n$, $n = 1...6$; basis $\{s, p_x, p_y\}$.

$$H = \begin{pmatrix} t_{ss\sigma}\sum_{n=1}^{6} e^{ik_n} & t_{sp\sigma}\left[e^{ik_1} - e^{ik_2} + \frac{1}{2}\left(e^{ik_3} - e^{ik_4} - e^{ik_5} + e^{ik_6}\right)\right] & \frac{\sqrt{3}}{2}t_{sp\sigma}\left(e^{ik_3} - e^{ik_4} + e^{ik_5} - e^{ik_6}\right) \\ & t_{pp\sigma}\sum_{n=1}^{2} e^{ik_n} + \frac{1}{4}\left(t_{pp\sigma} + 3t_{pp\pi}\right)\sum_{n=3}^{6} e^{ik_n} & \frac{\sqrt{3}}{4}\left(t_{pp\sigma} - t_{pp\pi}\right)\left(e^{ik_3} + e^{ik_4} - e^{ik_5} - e^{ik_6}\right) \\ \dagger & & t_{pp\pi}\sum_{n=1}^{2} e^{ik_n} + \frac{1}{4}\left(3t_{pp\sigma} + t_{pp\pi}\right)\sum_{n=3}^{6} e^{ik_n} \end{pmatrix}$$

$\downarrow$

$$H_{FB} = \begin{pmatrix} -\frac{2}{3}\sum_{n=1}^{6} e^{ik_n} & \sqrt{\frac{2}{3}}\left[e^{ik_1} - e^{ik_2} + \frac{1}{2}\left(e^{ik_3} - e^{ik_4} - e^{ik_5} + e^{ik_6}\right)\right] & \frac{1}{\sqrt{2}}\left(e^{ik_3} - e^{ik_4} + e^{ik_5} - e^{ik_6}\right) \\ & \sum_{n=1}^{2} e^{ik_n} & \frac{1}{\sqrt{3}}\left(e^{ik_3} + e^{ik_4} - e^{ik_5} - e^{ik_6}\right) \\ \dagger & & -\frac{1}{3}\sum_{n=1}^{2} e^{ik_n} + \frac{2}{3}\sum_{n=3}^{6} e^{ik_n} \end{pmatrix}$$

$\Gamma: \psi_{E=-4t} = |s\rangle,$
$\psi_{E=2t} = |p_x\rangle,$
$\psi_{E=2t} = |p_y\rangle$

## 3.2 $sd^2$-trigonal lattice: basis $\{s, d_{xy}, d_{x^2-y^2}\}$.

$$H = \begin{pmatrix} t_{ss\sigma}\sum_{n=1}^{6}e^{ik_n} & \frac{3}{4}t_{sd\sigma}\left(\sum_{n=3}^{4}e^{ik_n} - \sum_{n=5}^{6}e^{ik_n}\right) & \frac{\sqrt{3}}{2}t_{sd\sigma}\left(\sum_{n=1}^{2}e^{ik_n} - \frac{1}{2}\sum_{n=3}^{6}e^{ik_n}\right) \\ & t_{dd\pi}\sum_{n=1}^{2}e^{ik_n} + \frac{1}{16}(9t_{dd\sigma}+4t_{dd\pi}+3t_{dd\delta})\sum_{n=3}^{6}e^{ik_n} & \frac{\sqrt{3}}{16}(-3t_{dd\sigma}+4t_{dd\pi}-t_{dd\delta})\left(\sum_{n=3}^{4}e^{ik_n} - \sum_{n=5}^{6}e^{ik_n}\right) \\ & & \frac{1}{4}(3t_{dd\sigma}+t_{dd\delta})\sum_{n=1}^{2}e^{ik_n} + \frac{1}{16}(3t_{dd\sigma}+12t_{dd\pi}+t_{dd\delta})\sum_{n=3}^{6}e^{ik_n} \end{pmatrix}$$

$\downarrow$

$$H_{FB} = \begin{pmatrix} -\frac{2}{3}\sum_{n=1}^{6}e^{ik_n} & -\frac{1}{\sqrt{6}}\left(\sum_{n=3}^{4}e^{ik_n} - \sum_{n=5}^{6}e^{ik_n}\right) & -\frac{\sqrt{2}}{3}\left(\sum_{n=1}^{2}e^{ik_n} - \frac{1}{2}\sum_{n=3}^{6}e^{ik_n}\right) \\ & \sum_{n=1}^{2}e^{ik_n} & \frac{1}{\sqrt{3}}\left(\sum_{n=3}^{4}e^{ik_n} - \sum_{n=5}^{6}e^{ik_n}\right) \\ & & -\frac{1}{3}\sum_{n=1}^{2}e^{ik_n} + \frac{2}{3}\sum_{n=3}^{6}e^{ik_n} \end{pmatrix}$$

$\Gamma: \psi_{E=-4t} = |s\rangle,$
$\psi_{E=2t} = |d_{xy}\rangle,$
$\psi_{E=2t} = |d_{x^2-y^2}\rangle$

## 3.3 Kagome lattice: NN vectors $\boldsymbol{a}_1 = a(1,0)$, $\boldsymbol{a}_2 = \frac{a}{2}(1,\sqrt{3})$, $\boldsymbol{a}_3 = \boldsymbol{a}_2 - \boldsymbol{a}_1$; $k_n = \boldsymbol{k}\cdot\boldsymbol{a}_n$, $n=1,2,3$; basis $\{s_A, s_B, s_C\}$.

$$H = 2t^{NN}\begin{pmatrix} 0 & \cos k_1 & \cos k_2 \\ & 0 & \cos k_3 \\ \dagger & & 0 \end{pmatrix} \rightarrow H_{FB} = -2\begin{pmatrix} 0 & \cos k_1 & \cos k_2 \\ & 0 & \cos k_3 \\ \dagger & & 0 \end{pmatrix}$$

$\Gamma: \psi_{E=-4t} = \frac{1}{\sqrt{3}}(|A\rangle + |B\rangle + |C\rangle), \psi_{E=2t} = \frac{1}{\sqrt{2}}(-|A\rangle + |B\rangle),$
$\psi_{E=2t} = \frac{1}{\sqrt{6}}(-|A\rangle - |B\rangle + 2|C\rangle)$

**4.1 ($p_x$, $p_y$) hexagonal lattice**: NN vectors $a_1 = \dfrac{a}{2}(\sqrt{3},1)$, $a_2 = \dfrac{a}{2}(-\sqrt{3},1)$, $a_3 = -a_1 - a_2$; $k_n = \boldsymbol{k} \cdot \boldsymbol{a}_n$, $n=1,2,3$; basis $\{A{:}(p_x, p_y), B{:}(p_x, p_y)\}$.

$$H = \begin{pmatrix} 0 & 0 & \dfrac{1}{4}(3t_{pp\sigma} + t_{pp\pi})(e^{ik_1} + e^{ik_2}) + t_{pp\pi}e^{ik_3} & \dfrac{\sqrt{3}}{4}(t_{pp\sigma} - t_{pp\pi})(e^{ik_1} - e^{ik_2}) \\ 0 & \dfrac{\sqrt{3}}{4}(t_{pp\sigma} - t_{pp\pi})(e^{ik_1} - e^{ik_2}) & \dfrac{1}{4}(t_{pp\sigma} + 3t_{pp\pi})(e^{ik_1} + e^{ik_2}) + t_{pp\sigma}e^{ik_3} \\ & 0 & 0 \\ \dagger & & 0 \end{pmatrix}$$

↓

$$H_{FB} = \dfrac{1}{6}\begin{pmatrix} 0 & 0 & 3(e^{ik_1} + e^{ik_2}) & \sqrt{3}(e^{ik_1} - e^{ik_2}) \\ & 0 & \sqrt{3}(e^{ik_1} - e^{ik_2}) & (e^{ik_1} + e^{ik_2}) + 4e^{ik_3} \\ & & 0 & 0 \\ \dagger & & & 0 \end{pmatrix}$$

$\Gamma: \psi_{E=-t} = \dfrac{1}{\sqrt{2}}(-|A{:}p_x\rangle + |B{:}p_x\rangle), \psi_{E=-t} = \dfrac{1}{\sqrt{2}}(|A{:}p_y\rangle - |B{:}p_y\rangle)$

$\psi_{E=t} = \dfrac{1}{\sqrt{2}}(|A{:}p_x\rangle + |B{:}p_x\rangle), \psi_{E=t} = \dfrac{1}{\sqrt{2}}(|A{:}p_y\rangle + |B{:}p_y\rangle)$

**4.2 ($d_{xy}$, $d_{x2-y2}$) hexagonal lattice**: basis $\{A{:}(d_{xy}, d_{x^2-y^2}), B{:}(d_{xy}, d_{x^2-y^2})\}$.

$$H = \begin{pmatrix} 0 & 0 & \dfrac{1}{16}(9t_{dd\sigma} + 4t_{dd\pi} + 3t_{dd\delta})(e^{ik_1} + e^{ik_2}) + t_{dd\pi}e^{ik_3} & \dfrac{\sqrt{3}}{16}(3t_{dd\sigma} - 4t_{dd\pi} + t_{dd\delta})(e^{ik_1} - e^{ik_2}) \\ 0 & \dfrac{\sqrt{3}}{16}(3t_{dd\sigma} - 4t_{dd\pi} + t_{dd\delta})(e^{ik_1} - e^{ik_2}) & \dfrac{1}{16}(3t_{dd\sigma} - 12t_{dd\pi} + t_{dd\delta})(e^{ik_1} + e^{ik_2}) + \dfrac{1}{4}(3t_{dd\sigma} + t_{dd\delta})e^{ik_3} \\ & 0 & 0 \\ \dagger & & 0 \end{pmatrix}$$

$$H_{FB} = -H_{FB} \quad (4.1)$$

$$\Gamma: \psi_{E=-t} = -\frac{1}{\sqrt{2}}\left(\left|A:d_{xy}\right\rangle + \left|B:d_{xy}\right\rangle\right), \psi_{E=-t} = \frac{1}{\sqrt{2}}\left(\left|A:d_{x^2-y^2}\right\rangle + \left|B:d_{x^2-y^2}\right\rangle\right)$$

$$\psi_{E=t} = \frac{1}{\sqrt{2}}\left(\left|A:d_{xy}\right\rangle - \left|B:d_{xy}\right\rangle\right), \psi_{E=t} = \frac{1}{\sqrt{2}}\left(\left|A:d_{x^2-y^2}\right\rangle - \left|B:d_{x^2-y^2}\right\rangle\right)$$

**4.3 Diatomic-Kagome lattice**: basis $\{s_{A_1}, s_{A_2}, s_{A_3}, s_{B_1}, s_{B_2}, s_{B_3}\}$.

$$H = \begin{pmatrix} 0 & t^{NN}e^{ik_{A_1A_2}} & t^{NN}e^{ik_{A_1A_3}} & t^{NNN}e^{ik_{A_1B_1}} & 0 & 0 \\ & 0 & t^{NN}e^{ik_{A_2A_3}} & 0 & t^{NNN}e^{ik_{A_2B_2}} & 0 \\ & & 0 & 0 & 0 & t^{NNN}e^{ik_{A_3B_3}} \\ & & & 0 & t^{NN}e^{ik_{B_1B_2}} & t^{NN}e^{ik_{B_1B_3}} \\ & \dagger & & & 0 & t^{NN}e^{ik_{B_2B_3}} \\ & & & & & 0 \end{pmatrix} \rightarrow H_{FB} = -\begin{pmatrix} 0 & e^{ik_{A_1A_2}} & e^{ik_{A_1A_3}} & e^{ik_{A_1B_1}} & 0 & 0 \\ & 0 & e^{ik_{A_2A_3}} & 0 & e^{ik_{A_2B_2}} & 0 \\ & & 0 & 0 & 0 & e^{ik_{A_3B_3}} \\ & & & 0 & e^{ik_{B_1B_2}} & e^{ik_{B_1B_3}} \\ & \dagger & & & 0 & e^{ik_{B_2B_3}} \\ & & & & & 0 \end{pmatrix} \rightarrow$$

$$\Gamma: \psi_{E=-3t} = \frac{1}{\sqrt{6}}\begin{pmatrix} 1 \\ 1 \\ 1 \\ 1 \\ 1 \\ 1 \end{pmatrix}, \psi_{E=-t} = \frac{1}{\sqrt{6}}\begin{pmatrix} -1 \\ -1 \\ -1 \\ 1 \\ 1 \\ 1 \end{pmatrix}, \psi_{E=0} = \frac{1}{2}\begin{pmatrix} -1 \\ 1 \\ 0 \\ -1 \\ 1 \\ 0 \end{pmatrix}, \psi_{E=0} = \frac{1}{2\sqrt{3}}\begin{pmatrix} 1 \\ 1 \\ -2 \\ 1 \\ 1 \\ -2 \end{pmatrix}, \psi_{E=2t} = \frac{1}{2}\begin{pmatrix} 1 \\ -1 \\ 0 \\ -1 \\ 1 \\ 0 \end{pmatrix}, \psi_{E=2t} = \frac{1}{2\sqrt{3}}\begin{pmatrix} 1 \\ 1 \\ -2 \\ -1 \\ -1 \\ 2 \end{pmatrix}$$